\documentclass[english,preprint]{elsarticle}
\usepackage[T1]{fontenc}
\usepackage[latin9]{inputenc}
\usepackage{booktabs}
\usepackage{amsthm}
\usepackage{amsmath}
\usepackage{amssymb}
\usepackage{wasysym}
\usepackage{graphicx}

\makeatletter

\newcommand{\noun}[1]{\textsc{#1}}
\providecommand{\tabularnewline}{\\}

\theoremstyle{plain}
\newtheorem{thm}{\protect\theoremname}
  \theoremstyle{plain}
  \newtheorem{cor}[thm]{\protect\corollaryname}

\@ifundefined{definecolor}
 {\usepackage{color}}{}
\usepackage{amsthm}\usepackage{wasysym}

\usepackage{booktabs}\newcommand{\ra}[1]{\renewcommand{\arraystretch}{#1}}
\ra{1.3}

\journal{Discrete and Applied Mathematics}

\makeatother

\usepackage{babel}
  \providecommand{\corollaryname}{Corollary}
\providecommand{\theoremname}{Theorem}

\begin{document}

\title{Simple Agents Learn to Find Their Way:\\
An Introduction on Mapping Polygons}

\author[LIF]{Jérémie Chalopin}

\ead{jeremie.chalopin@lif.univ-mrs.fr}

\address[LIF]{LIF, CNRS \& Aix-Marseille Université, France}

\author[TEC]{Shantanu Das}

\ead{shantanu@tx.technion.ac.il}

\address[TEC]{BGU \& Technion -- Israel Institute of Technology, Israel}

\author[ETH]{Yann Disser\corref{cor}}

\ead{ydisser@inf.ethz.ch}

\cortext[cor]{Corresponding author}

\author[ETH]{Matúš Mihalák}

\ead{mmihalak@inf.ethz.ch}

\author[ETH]{Peter Widmayer}

\ead{widmayer@inf.ethz.ch}

\address[ETH]{Institute of Theoretical Computer Science, ETH Zürich, Switzerland}
\begin{abstract}
This paper gives an introduction to the problem of mapping simple
polygons with autonomous agents. We focus on minimalistic agents that
move from vertex to vertex along straight lines inside a polygon,
using their sensors to gather local observations at each vertex. Our
attention revolves around the question whether a given configuration
of sensors and movement capabilities of the agents allows them to
capture enough data in order to draw conclusions regarding the global
layout of the polygon.

In particular, we study the problem of reconstructing the visibility
graph of a simple polygon by an agent moving either inside or on the
boundary of the polygon. Our aim is to provide insight about the algorithmic
challenges faced by an agent trying to map a polygon. We present an
overview of techniques for solving this problem with agents that are
equipped with simple sensorial capabilities. We illustrate these techniques
on examples with sensors that measure angles between lines of sight
or identify the previous location. We give an overview over related
problems in combinatorial geometry as well as graph exploration.\end{abstract}
\maketitle

\section{Introduction}

There is a continuing trend in robotics towards designing and building
simple and cheap microrobots that collaborate to perform certain tasks,
rather than using a single, expensive and complicated robot. Typical
examples include groups of microrobots (such as simple sensors and
actuators, simple wireless sensors, simple ``insect''-type microrobots)
that are employed to collectively guard an area, or to collectively
sweep an area in order to find a potential intruder. Reasons in favor
of microrobots are not only cost savings, but also limitations imposed
by the environment, conditions suggested by the task at hand (which
restrict the usage of rather bulky robots), and, last but not least,
robustness considerations: If a few microrobots in a swarm turn out
not to be operational or get damaged over time, the rest of the crowd
might still be able to complete the task successfully.

This microrobot design trend has inspired the question what a group
of simple microrobotic agents or a single such agent can or cannot
do, given limitations on the agents' abilities. These limitations
affect what a microrobot can sense, how a microrobot can move, and
how microrobots can communicate, in addition to computational and
memory limitations. We aim at understanding which abilities are indispensable
for a microrobot to perform a given task. In other words, we aim to
identify minimal sets of abilities such that these abilities allow
the robot to perform the task at hand, but any weaker configuration
is insufficient. This leads to very simple, abstract robots that we
will call ``agents'' from now on. Just like Columbus could not identify
what he saw when he landed in America, the agents we are going to
use throughout the next sections have no way of directly identifying
the objects they sense. They are worse off than Columbus in that they
have no notion of coordinates or of distances. We pose the most basic
questions in such a setting: What can our agents learn about their
initially unknown environment? Can they navigate in it? Can they join
forces to achieve a common goal? What goals are impossible for them
to reach?

Over the past decade, a wealth of fundamental research results have
identified a variety of very limited configurations of abilities that
still allow interesting tasks to be performed. This paper aims to
serve as an introduction for newcomers to this field. The exposition
is guided by the authors' own investigations, and in the core part
of the paper it is limited to these. We describe an extremely simple
kind of agents, working in a very simple environment, confronted with
the task of inferring the layout of the environment. Specifically,
most of the paper is concerned with a single agent that operates inside
a polygon in the plane and needs to construct a ``map'' of the polygon.
Our main goal is to understand for which settings a single agent
can acquire enough pieces of \emph{local} information to finally infer
a \emph{global} picture of its environment. We present how, in some
cases, certain local geometric information alone can be used to obtain
a global picture. In other, more involved cases, we learn how concepts
from distributed computing theory (even for a single agent) help,
and how ideas from the general graph-exploration setting carry over
to the geometric setting of polygon exploration. We also show how
the stronger geometrical structure helps to achieve goals that cannot
be achieved in general graphs.

Rather than confronting the reader with an abstract top-down coverage
of the field, we try to delve into the world of polygon exploration
with as little ado as possible. To that end we focus on a few characteristic
techniques for mapping polygons in detail, and explain them using
many illustrations and examples instead of a formal approach. A lightweight
formal foundation is provided in Section~\ref{sec:model}. After
that, we begin the tour in Section~\ref{sec:boundary_only} considering
agents that can measure angles induced by any two visible vertices.
In Section~\ref{sec:base_graph}, we shift our attention to agents
that can move across the polygon along straight lines between vertices,
but on the other hand have weaker sensors. With Section~\ref{sec:summary_mapping}
we finish the discussion about polygon exploration by providing a
list of other known results in the area, briefly illustrating the
different agent models that have been studied. Finally, Section~\ref{sec:related_work}
offers a glimpse on related problems beyond mapping, beyond agents,
and beyond polygonal environments.

More details on some of the results presented in this paper, can be
found in~\citep{DisserPhD/11}.

\section{The visibility graph reconstruction problem -- model and notation\label{sec:model}}

\begin{figure}
\begin{centering}
\includegraphics[width=0.8\columnwidth]{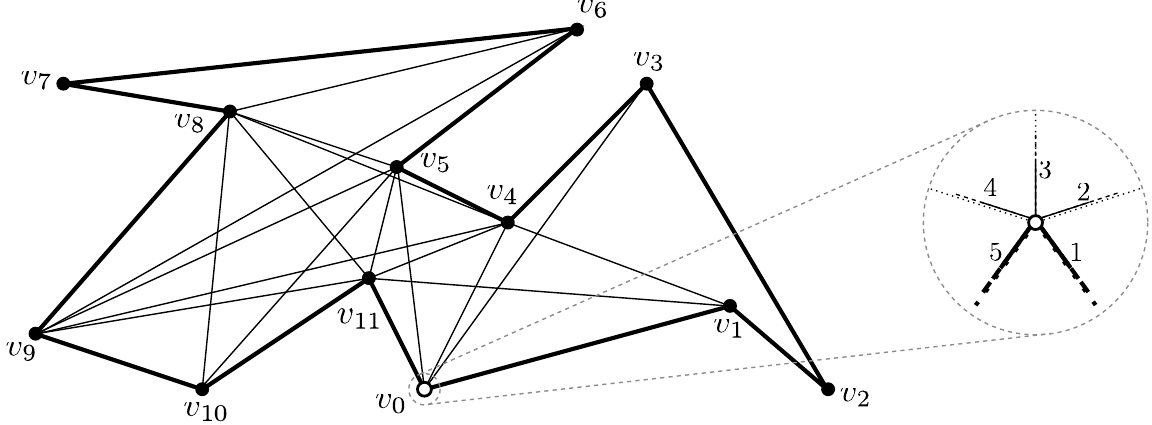} 
\par\end{centering}

\caption{Left: A simple polygon with embedded visibility graph. Right: The
polygon as perceived locally by an agent located at $v_{0}$. All
that the agent can observe is an order of the edges incident to its
location. Except for the two boundary-edges, the agent does not know
to which vertex, or in which direction, each edge leads.\label{fig:Visibility-graph+agent}}
\end{figure}

In the following, we consider a mobile agent exploring a simple polygon
$\mathcal{P}$ with $n$ vertices (cf.~Figure~\ref{fig:Visibility-graph+agent}).
The goal of the agent is to reconstruct the visibility graph $G_{\mathrm{vis}}$
of $\mathcal{P}$, i.e., the graph having a node for every vertex
of the polygon, and an edge connecting two nodes if the corresponding
vertices see each other. Two vertices of $\mathcal{P}$ are said to
see each other if and only if the line segment connecting them lies
in $\mathcal{P}$ entirely. In that way, every edge of $G_{\mathrm{vis}}$
corresponds to a line segment in $\mathcal{P}$, and we refer to edges
and the corresponding line segments interchangeably. Similarly, we
identify each vertex of $\mathcal{P}$ with the corresponding node
of $G_{\mathrm{vis}}$. We fix some vertex $v_{0}$ and write $v_{0},v_{1},\ldots,v_{n-1}$
to denote the vertices of $\mathcal{P}$ in the order that they are
encountered along a tour of the boundary which starts at $v_{0}$
and locally has the interior of $\mathcal{P}$ on its left. From now
on, we refer to this order as the counter-clockwise order along the
boundary of $\mathcal{P}$. Finally, we write $d_{i},0\leq i<n$,
to denote the degree of $v_{i}$, i.e., the number of edges incident
to $v_{i}$ in $G_{\mathrm{vis}}$. For convenience, all operations
on indices are understood modulo $n$.

This paper is centered around the question how much local data has
to be collected about $\mathcal{P}$ in order to infer $G_{\mathrm{vis}}$.
In order to capture the data collection in $\mathcal{P}$ formally,
we introduce a very basic agent that moves inside $\mathcal{P}$ and
makes local observations in the process. The idea is to keep the agent
model as simplistic as possible and extend it later depending on what
local data we are interested in. The agent is modeled as a point (a
dimensionless object) moving from vertex to vertex along straight
lines inside $\mathcal{P}$, i.e., along edges of $G_{\mathrm{vis}}$.
In the beginning, the agent is located at $v_{0}$, and the only information
it has about $\mathcal{P}$ is that $\mathcal{P}$ is a simple polygon.
The agent does not know any information about $n$, the number of
vertices of $\mathcal{P}$. While located at a vertex $v_{i}$, the
agent perceives the edges of $G_{\mathrm{vis}}$ incident to $v_{i}$
in counter-clockwise order, starting with the boundary edge $\left(v_{i},v_{i+1}\right)$
(cf.~Figure~\ref{fig:Visibility-graph+agent}). We choose $v_{0}$
to be the agent's initial location, therefore the agent can keep track
of its global position as long as it moves along the boundary only.
However, it can neither perceive the global index $i$ of its location
$v_{i}$ directly, nor the global indices of the vertices to which
the edges from $v_{i}$ lead. This means that once it moves along
an edge through the inside of $\mathcal{P}$, in a way, the agent
looses sense of its global position. The counter-clockwise ordering
of the edges at a vertex is the only means of orientation that the
agent has when deciding a move. The move itself is assumed to be instantaneous,
i.e., the agent cannot make any observations while moving. Every
movement decision of the agent and the conclusions it draws from local
observations are based on all the information it has collected so
far -- a history of movement decisions and observed vertex degrees.
Because our focus is to study the effect of movement and sensing capabilities,
we do not restrict the agent computationally, and we assume that the
agent has enough memory to store all the history of movements and
observations. The question is whether the information collected this
way is sufficient for the agent to infer $G_{\mathrm{vis}}$.

From now on, we will refer to an agent in the above model as the \emph{basic
agent. }In later sections, variants of the basic agent will be introduced,
mainly extending its sensing capabilities. For a particular extension
of the model, the central question is whether the agent can solve
the \emph{visibility graph reconstruction problem}, i.e., whether
it can infer $G_{\mathrm{vis}}$ (up to renaming of the nodes) within
a finite number of operations. If this is possible, we say that the
agent \emph{can reconstruct $G_{\mathrm{vis}}$. }We will sometimes
use the expression of \emph{constructing} a visibility graph $G_{\mathrm{vis}}^{\prime}$
that fits the observations of the agent, in the sense that there is
a polygon $\mathcal{P}^{\prime}$ with visibility graph $G_{\mathrm{vis}}^{\prime}$,
such that the agent would make the same observations in $\mathcal{P}^{\prime}$
as in $\mathcal{P}$. In contrast, the \emph{reconstruction} problem
requires the agent to be able to find $G_{\mathrm{vis}}$ itself.
Our challenge is to obtain frugal models which already enable the
agent to solve this visibility graph reconstruction problem in finite
time.

\begin{figure}
\begin{centering}
\includegraphics[width=1\columnwidth]{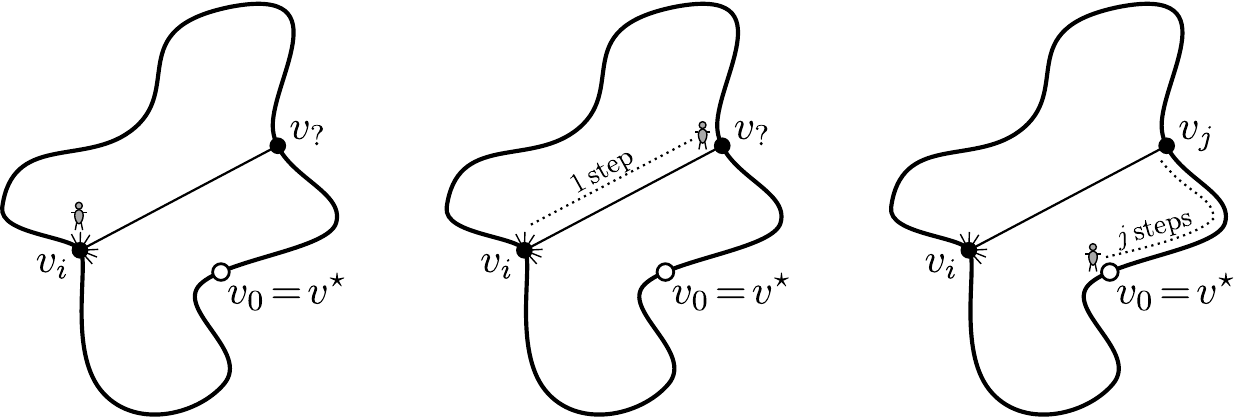} 
\par\end{centering}

\caption{If the agent can distinguish some vertex $v^{\star}$ (in our example
$v^{\star}=v_{0}$), it can easily determine where edges lead: starting
at a vertex $v_{i}$, the agent can identify the target of an edge
by moving along this edge and then along the boundary until it encounters
$v^{\star}$, counting the number of moves it makes.\label{fig:distinguish-vertex}}
\end{figure}

In order to reconstruct the visibility graph, it is sufficient to
decide for every vertex where the edges incident to this vertex lead
in terms of global identities (i.e., global indices). This task becomes
trivial if there is a vertex $v^{\star}$ with the property that the
agent can distinguish at any time whether or not it is currently located
at $v^{\star}$. In that case, the agent can decide where an edge
leads simply by moving along the edge and then counting the number
of moves along the boundary that it takes to get back to $v^{\star}$
(cf.~Figure~\ref{fig:distinguish-vertex}). Hence, the visibility
graph reconstruction problem is non-trivial only if no individual
vertex of $\mathcal{P}$ can be recognized by the agent. In some sense,
the problem is difficult only if $\mathcal{P}$ is symmetric with
respect to the data which the agent is able to perceive.

In the next two sections we will demonstrate two main solution approaches
that proved to be successful for analysing the visibility graph reconstruction
problem. For the first approach, we show how arguments from computational
geometry can be applied to our problem when the agent's sensing capabilities
provide enough geometric information about the polygon. In Section
\ref{sec:boundary_only}, we consider an agent which can measure the
angles between the lines from its current location to all visible
vertices. We will see that this ability provides enough geometric
information to reconstruct the shape of the polygon (and thus also
the visibility graph of the polygon). In this approach, the difficulty
lies in using the geometric information the sensing provides. For
the second approach, we show how ideas from distributed computing
apply in the geometric setting. In Section \ref{sec:base_graph} we
will demonstrate an approach in which the movements of the agent play
a key role. In that setting, the agent first obtains a rough map of
the environment by moving around in a systematic fashion. Intuitively,
the agent computes an arc-labeled directed graph in which a node represents
a set of vertices of the polygon that appear indistinguishable to
the agent, and the labels of the outgoing arcs encode the information
that the robot senses at a vertex. This rough map and the specific
geometric interpretation of the edge-labels is then used to reconstruct
the visibility graph of the polygon.

\section{Reconstruction from geometrical data\label{sec:boundary_only}}

In this section we consider the \emph{angle agent} that, roughly speaking,
is able to measure the exact angle between any two lines to vertices
visible from its current location. Based on the results in~\citep{DisserMihalakWidmayer/11},
we will show that an angle agent can reconstruct the visibility graph.
To present the key ideas of the reconstruction method, we will assume
that an angle agent knows $n$, the number of vertices of the polygon.
It can be shown, however, that the knowledge of $n$ is not necessary,
and that in fact the angle agent can infer $n$ \citep{DisserMihalakWidmayer/10b}.

\begin{figure}
\centering{}\includegraphics[width=0.9\columnwidth]{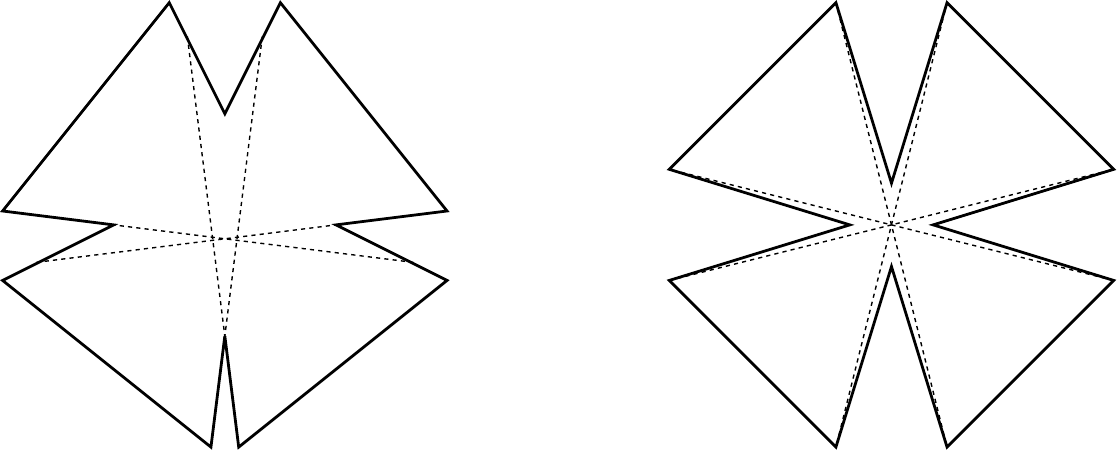}\caption{Two polygons with different visibility graphs that yield the same
observations to a basic agent that moves once around the boundary.\label{fig:no sensor counterex}}
\end{figure}

Recall that the basic agent is allowed to freely move along the edges
of $G_{\mathrm{vis}}$. In constrast, it will turn out that the angle
agent can reconstruct the visibility graph by moving along the edges
of the boundary only. Because the agent does not need to use all of
its movement capabilities, in a way, the ability to measure angles
is a ``strong'' ability. In Section~\ref{sec:base_graph} we will
see a ``weaker'' agent which extensively uses its movement capabilities
in order to reconstruct the visibility graph. It is worth mentioning
that the basic agent is not able to reconstruct the visibility graph
by moving along the boundary only, as Figure~\ref{fig:no sensor counterex}
illustrates.

\begin{figure}
\centering{}\hspace*{\fill}\includegraphics[width=0.4\columnwidth]{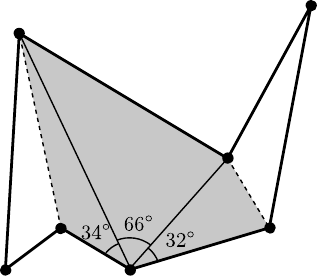}\hspace*{\fill}\includegraphics[width=0.35\columnwidth]{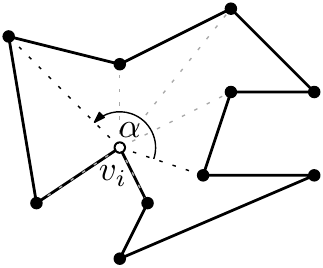}\hspace*{\fill}\caption{Left: An angle measurement at a vertex yields the list of angles between
adjacent edges of $G_{\mathrm{vis}}$ in counter-clockwise order,
here $\left(32^{\circ},66^{\circ},34^{\circ}\right)$. Right: The
angle between non-adjacent edges can easily be computed by summing
the angles in between.\label{fig:angles--angle_measurement}}
\end{figure}

We now define the angle agent more precisely as an extension of the
basic agent. In addition to the abilities of the basic agent, the
angle agent can perform, before each move, an \emph{angle measurement
}at the vertex $v_{i}$ of $\mathcal{P}$ that it is currently located
at. For every two vertices $u,w$ visible to $v_{i}$ (in this order),
the angle measurement returns the counter-clockwise angle formed by
the line-segments $\overline{v_{i}u}$ and $\overline{v_{i}w}$ (in
this order). We call this angle the \emph{angle at $v_{i}$ between
$u$ and $w$, }or simply the \emph{angle between $u$ and $w$ }if
the location of the agent is clear from the context. By $\measuredangle_{v}\!\left(u,w\right)$
we denote the counter-clockwise angle between the line segments $\overline{vu}$
and $\overline{vw}$ in this order, even if the vertices involved
do not see each other mutually. An angle measurement at vertex $v_{i}$
can be represented by a list $\left(\alpha_{1},\alpha_{2},\ldots,\alpha_{d_{i}-1}\right)$,
where $d_{i}$ is the degree of $v_{i}$, and $\alpha_{j}$ denotes
the angle at $v_{i}$ between the $j$-th and the $\left(j+1\right)$-th
vertex visible to $v_{i}$ (cf.~Figure~\ref{fig:angles--angle_measurement}).
Observe that we can represent the angle measurement by this list of
length $d_{i}-1$, as the angle between any two vertices $u$, $v$
visible to $v_{i}$ is easy to infer by summing the enclosed angles
(cf. Figure \ref{fig:angles--angle_measurement}).

We note that if the angle agent can reconstruct the visibility graph
of the underlying polygon $\mathcal{P}$, then it can also reconstruct
the geometry of the polygon up to similarity (translation, rotation,
and scaling): From $G_{\mathrm{vis}}$, the agent can compute a triangulation
of $\mathcal{P}$ and fix the positions of $v_{0}$ and $v_{1}$ in
the plane. As the angle agent can infer all three angles of each triangle
in the triangulation from the angle measurements, the position of
two vertices of a triangle in the plane is sufficient to derive the
position of the third. The agent can therefore gradually deduce the
positions of all vertices by repeatedly considering triangles where
exactly two positions are known, starting with the triangle containing
both $v_{0}$ and $v_{1}$. As a consequence, the visibility graph
reconstruction problem can equivalently be seen as the problem of
reconstructing the geometry of $\mathcal{P}$ (up to similarity).

\begin{figure}
\centering{}\includegraphics[scale=1.1]{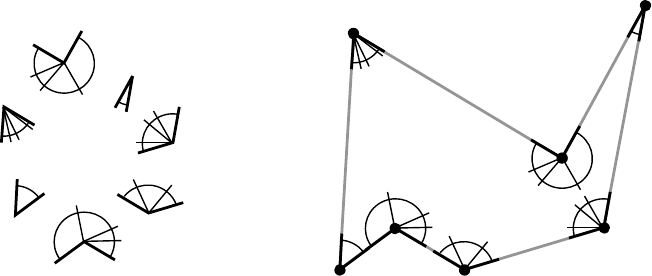}
\caption{Given the angle measurement for each vertex in counter-clockwise order
(left), the goal is to find the unique polygon that fits these angle
measurements (right). \label{fig:angles--problem_illustration}}
\end{figure}

As we will see, the angle agent is able to solve the visibility graph
reconstruction problem from the data collected during a single tour
of the boundary \citep{DisserMihalakWidmayer/11}. In this case, the
visibility graph reconstruction problem can be reformulated as a purely
geometrical problem that does not involve a mobile agent at all: We
can forget about the agent once it has moved once fully around $\mathcal{P}$,
gathering the angle measurements at every vertex. Note that here the
agent needs its knowledge about $n$ -- without, it is not clear how
far to move in order to have traveled once around the boundary. After
its tour of the boundary, the agent has gathered a list of the angle
measurements at every vertex, ordered as they appear along the boundary
in counter-clockwise order. We call this the \emph{ordered list of
angle measurements.} The difficulty of the reconstruction problem
entirely lies in how to use this data algorithmically. The reconstruction
problem becomes (cf.~Figure~\ref{fig:angles--problem_illustration})

\begin{quote}
\emph{Given the ordered list of angle measurements of $\mathcal{P}$,
find $G_{\mathrm{vis}}$.}
\end{quote}

We note that with weaker sensors a single tour along the boundary
does not suffice to reconstruct the visibility graph, and that a more
involved movement and data collection strategy is necessary (as we
will see in Section~\ref{sec:base_graph}). In turn, the problem
of reconstructing a visibility graph cannot in general be formulated
as a purely geometrical problem.

It is worth stressing that the reconstruction problem asks for the
unique visibility graph $G_{\mathrm{vis}}$ of $\mathcal{P}$, and
hence the shape of $\mathcal{P}$. We are not content with just \emph{some}
polygon $\mathcal{P}^{\prime}$ that has the same ordered list of
angle measurements. One can easily find some polygon that fits the
angle measurements by trying all possible visibility graphs with $n$
vertices, until a compatible one is found. On the other hand, it is
not clear whether we can always deduce $\mathcal{P}$ itself and guarantee
that we really reconstructed the original polygon (up to similarity),
and not some other polygon with the same ordered list of angle measurements.
In other words: Does an ordered list of angle measurements uniquely
determine a polygon? We will show that this is the case.

\subsection{A reconstruction algorithm\label{sec:angles-solution}}

We prove that $G_{\mathrm{vis}}$ can be reconstructed from the ordered
list of angle measurements of $\mathcal{P}$. We do this by developing
an iterative algorithm that constructs a compatible visibility graph.
We will show that this algorithm is guaranteed to reconstruct the
original visibility graph $G_{\mathrm{vis}}$.

\begin{figure}[t]
\centering{}\includegraphics{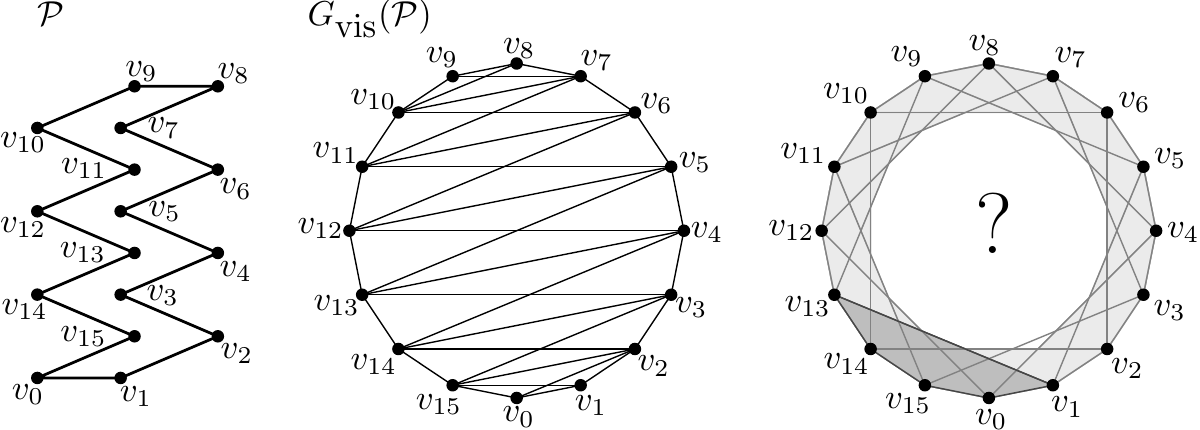} \caption{Progress of the reconstruction of a polygon $\mathcal{P}$ after step
$k=4$. From left to right: the polygon $\mathcal{P}$, its visibility
graph $G_{\mathrm{vis}}$, and the subgraphs of $G_{\mathrm{vis}}$
that have fully been reconstructed after $k=4$ steps (schematically
depicted as shaded areas -- no edges of the visibility graph are depicted).
All subgraphs of $G_{\mathrm{vis}}$ induced by $k+1=5$ consecutive
vertices (shaded areas) are already fully reconstructed. Edges between
vertices that are more than $k$ steps apart along the boundary have
not been reconstructed yet.\label{fig:angles--step_k_general}}
\end{figure}

The algorithm gradually builds $G_{\mathrm{vis}}=\left(V,E\right)$.
In the beginning, we start with a graph $G^{\left(0\right)}=(V,E^{\left(0\right)})$
with an empty set of edges $E^{\left(0\right)}=\emptyset$. In step
$k=1,2,\ldots,\lceil\frac{n}{2}\rceil$ we build $G^{\left(k\right)}=(V,E^{\left(k\right)})$
by identifying all edges of $G_{\mathrm{vis}}$ of the form $\left\{ v_{i},v_{i+k}\right\} $
and adding them to $E^{\left(k-1\right)}$ in order to obtain $E^{\left(k\right)}$.
In other words, $G^{\left(k\right)}$ contains exactly the edges of
$G_{\mathrm{vis}}$ that connect vertices which are at most $k$ edges
apart along the boundary (cf.~Figure~\ref{fig:angles--step_k_general}).
By definition, every subgraph of $G^{\left(k\right)}$ induced by
at most $k+1$ consecutive vertices along the boundary is equal to
the subgraph of $G_{\mathrm{vis}}$ induced by the same vertices.
Observe that for $k\geq\lceil\frac{n}{2}\rceil$, we have $G^{\left(k\right)}=G_{\mathrm{vis}}$
as no two vertices can be further apart than $\lceil\frac{n}{2}\rceil$
edges along the boundary. The only ingredient we are missing for a
complete algorithm is a criterion to decide in step $k$, for each
$v_{i}\in V$, whether $\left\{ v_{i},v_{i+k}\right\} \in E$. If
we can find a criterion that is both necessary and sufficient, we
obtain an algorithm that is guaranteed to reconstruct the visibility
graph $G_{\mathrm{vis}}$.

\begin{figure}[t]
\centering{}\includegraphics{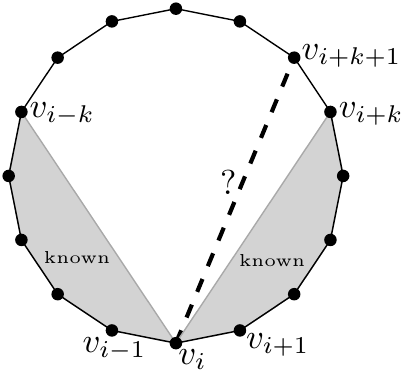} \caption{State of the reconstruction of $G_{\mathrm{vis}}$ after step $k$
of the algorithm. For every vertex $v_{i}$, the neighbors of $v_{i}$
in $\{v_{i+1},\ldots,v_{k}\}$ have already been identified. Hence,
the subgraphs of $G_{\mathrm{vis}}$ depicted by the shaded areas
have been fully reconstructed. In the next step, the algorithm has
to decide whether $v_{i}$ sees $v_{i+k+1}$, and thus whether $\{v_{i},v_{i+k+1}\}$
is an edge of $G_{\mathrm{vis}}$.\label{fig:angles--step_k}}
\end{figure}

Clearly, for $k=1$, the criterion is trivial: every vertex sees both
its neighbors along the boundary, and thus every edge of the form
$\left\{ v_{i},v_{i+1}\right\} $ is part of $G_{\mathrm{vis}}$ and
needs to be added to $E^{\left(1\right)}$. By induction, it remains
to be shown how to decide whether $\left\{ v_{i},v_{i+k+1}\right\} \in E$
in step $k+1$ of the algorithm, assuming that $G^{\left(k\right)}$
has already been computed. As $G^{\left(k\right)}$ is known, we know
for every vertex $v_{i}\in V$ which vertices in $\left\{ v_{i+1},v_{i+2},\ldots,v_{i+k}\right\} $
and in $\left\{ v_{i-k},v_{i-k+1},\dots,v_{i-1}\right\} $ are visible
to $v_{i}$ (cf.~Figure~\ref{fig:angles--step_k}). The remainder
of this section shows how to use this information in order to decide
whether $\left\{ v_{i},v_{i+k+1}\right\} \in E$ for any $v_{i}\in V$.

\begin{figure}
\centering{}\hspace*{\fill}\includegraphics[width=0.45\columnwidth]{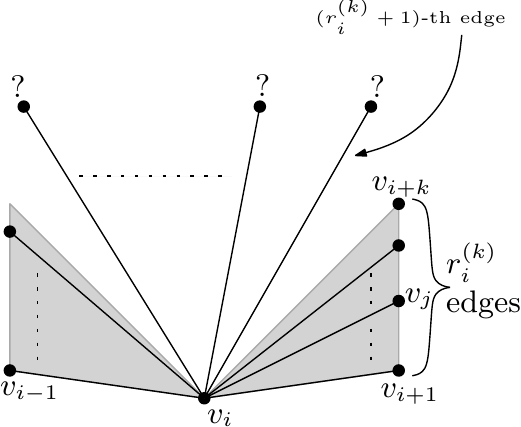}\hspace*{\fill}\includegraphics[width=0.45\columnwidth]{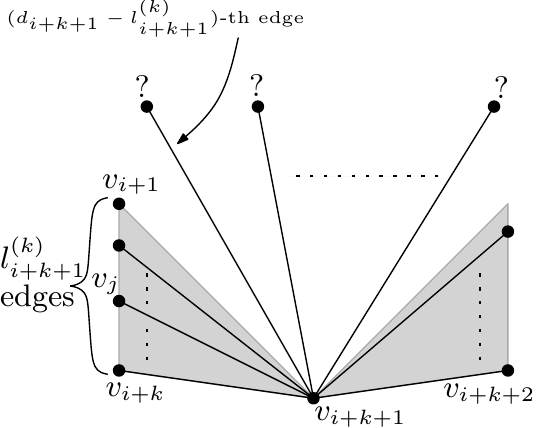}\hspace*{\fill}\caption{Knowledge about the edges of $v_{i}$ and $v_{i+k+1}$ after step
$k$ of the algorithm. Left: the edges from $v_{i}$ to vertices in
$\{v_{i+1},\ldots,v_{i+k}\}$ and in $\{v_{i-1},\ldots,v_{i-k}\}$
have been identified. If $v_{i}$ has $r_{i}^{(k)}$ neighbors in
$\{v_{i+1},\ldots,v_{i+k}\}$, then deciding whether $v_{i}$ sees
$v_{i+k+1}$ is equivalent to deciding whether the $(r_{i}^{\left(k\right)}+1)$-th
edge of $v_{i}$ leads to $v_{i+k+1}$. Right: If $v_{i+k+1}$ has
$l_{i+k+1}^{\left(k\right)}$ neighbors in $v_{i+1},\ldots,v_{i+k}$,
then deciding whether $v_{i+k+1}$ sees $v_{i}$ is equivalent to
deciding whether the $(d_{i+k+1}-l_{i+k+1}^{\left(k\right)})$-th
edge of $v_{i+k+1}$ leads to $v_{i}$.\label{fig:angles--some_vertices_known}}
\end{figure}

By $r_{i}^{\left(k\right)}$ we denote the number of vertices that
$v_{i}$ sees on its ``right'' among $\left\{ v_{i+1},v_{i+2},\ldots,v_{i+k}\right\} $,
i.e., $r_{i}^{\left(k\right)}$ is the degree of $v_{i}$ in the subgraph
of $G_{\mathrm{vis}}$ induced by the set of vertices $\left\{ v_{i},v_{i+1},\ldots,v_{i+k}\right\} $.
Similarly, $l_{i}^{\left(k\right)}$ is the degree of $v_{i}$ in
the subgraph of $G_{\mathrm{vis}}$ induced by the set of vertices
$\left\{ v_{i-k},v_{i-k+1},\ldots,v_{i}\right\} $. If $v_{i}$ sees
$v_{i+k+1}$, then the corresponding edge is the $(r_{i}^{\left(k\right)}+1)$-th
edge of $v_{i}$ and the $(d_{i+k+1}-l_{i+k+1}^{\left(k\right)})$-th
edge of $v_{i+k+1}$ (cf.~Figure~\ref{fig:angles--some_vertices_known}).
Recall that the edges are locally ordered in counter-clockwise order.

\begin{figure}[t]
\centering{}\includegraphics[scale=0.75]{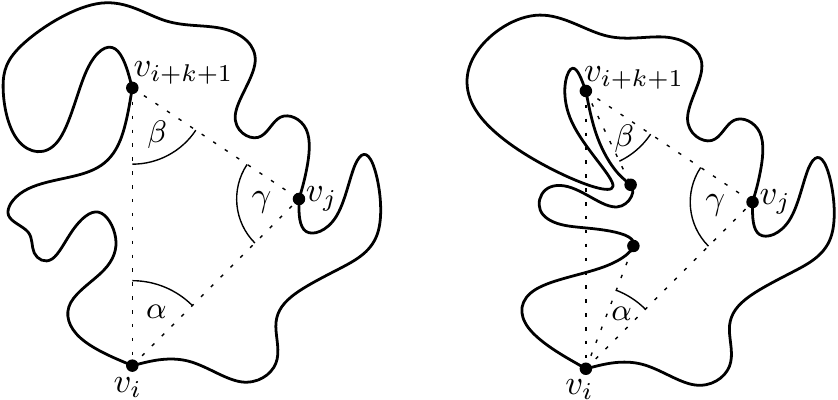}
\caption{Schematic illustration of the induction step of the reconstruction
of $G_{\mathrm{vis}}$. Left: If $v_{i}$ sees $v_{i+k+1}$, there
has to be a vertex $v_{j}$ in $\left\{ v_{i+1},\ldots,v_{i+k}\right\} $
which sees both $v_{i}$ and $v_{i+k+1}$, and $v_{i},v_{j},v_{i+k+1}$
form a triangle in $G_{\mathrm{vis}}$. Right: If $v_{i}$ does not
see $v_{i+k+1}$, and $v_{j}$ sees $v_{i}$ and $v_{i+k+1}$at an
angle $\gamma<\pi$, the boundary of $\mathcal{P}$ between $v_{i+k+1}$
and $v_{i}$ (in this order) has to block the line of sight between
$v_{i}$ and $v_{i+k+1}$.\label{fig:angles--induction_step}}
\end{figure}

Let us first consider the case in which $v_{i}$ does see $v_{i+k+1}$.
Then, we claim, there has to be a vertex in $\left\{ v_{i+1},v_{i+2},\ldots,v_{i+k}\right\} $
which sees both $v_{i}$ and $v_{i+k+1}$. To see this, observe that
there is a triangulation of $\mathcal{P}$ which uses the edge $\left\{ v_{i},v_{i+k+1}\right\} $
in two triangles. As $k\geq1$, there must be a vertex $v_{j}\in\{v_{i+1},v_{i+2},$
$\ldots,v_{i+k}\}$ that forms a triangle in the thriangulation with
$v_{i}$ and $v_{i+k+1}$. By definition, $v_{j}$ sees both $v_{i}$
and $v_{i+k+1}$ (cf.~Figure~\ref{fig:angles--induction_step}),
which proves our claim. Evidently, for every $v_{j}\in\left\{ v_{i+1},v_{i+2},\ldots,v_{i+k}\right\} $
which sees both $v_{i}$ and $v_{i+k+1}$, we have that $\measuredangle_{v_{i}}\!\left(v_{j},v_{i+k+1}\right)+\measuredangle_{v_{j}}\!\left(v_{i+k+1},v_{i}\right)+\measuredangle_{v_{i+k+1}}\!\left(v_{i},v_{j}\right)=\pi$.

Based on the above, we can derive a checkable necessary condition
which has to hold if $v_{i}$ sees $v_{i+k+1}$. First of all, in
the set $\left\{ v_{i+1},v_{i+2},\ldots,v_{i+k}\right\} $ there has
to be a vertex which sees both $v_{i}$ and $v_{i+k+1}$. By induction,
$G^{\left(k\right)}$ has already been computed and therefore we can
easily check whether such a vertex $v_{j}$ exists. Now let $x,y$
be such that $v_{j}$ is the $x$-th vertex visible to $v_{i}$ and
the $y$-th vertex visible to $v_{i+k+1}$. We can look up the angle
$\alpha$ at $v_{i}$ between the $x$-th and the $(r_{i}^{\left(k\right)}+1)$-th
edge and the angle $\beta$ at $v_{i+k+1}$ between the $(d_{i+k+1}-l_{i+k+1}^{\left(k\right)})$-th
and the $y$-th edge in the angle measurements at $v_{i}$ and $v_{i+k+1}$,
respectively. Similarly, we can look up the angle $\gamma=\measuredangle_{v_{j}}\!\left(v_{i+k+1},v_{i}\right)$
as we know which edges at $v_{j}$ lead to $v_{i}$ and to $v_{i+k+1}$.
If $v_{i}$ sees $v_{i+k+1}$, we must have $\alpha+\beta+\gamma=\pi$
as then $\alpha=\measuredangle_{v_{i}}\!\left(v_{j},v_{i+k+1}\right)$
and $\beta=\measuredangle_{v_{i+k+1}}\!\left(v_{i},v_{j}\right)$.
In total, we get that if $v_{i}$ sees $v_{i+k+1}$, we can find a
vertex in $\left\{ v_{i+1},v_{i+2},\ldots,v_{i+k}\right\} $ seeing
both $v_{i}$ and $v_{i+k+1}$, and for each such vertex $v_{j}$
we must have $\alpha+\beta+\gamma=\pi$. We will argue that this condition
is also sufficient for $v_{i}$ and $v_{i+k+1}$ to see each other.
As the condition is easy to check, this completes the inductive step
of the algorithm.

In order to prove that our condition is sufficient, we need to show
that if $v_{i}$ does not see $v_{i+k+1}$, then for every vertex
$v_{j}$ in $\left\{ v_{i+1},v_{i+2},\ldots,v_{i+k}\right\} $ that
sees both $v_{i}$ and $v_{i+k+1}$, we have $\alpha+\beta+\gamma\neq\pi$.
This is immediate for all such $v_{j}$'s for which $\gamma>\pi$,
so it remains to consider $v_{j}$'s for which $\gamma\leq\pi$. Observe
that, as $v_{i}$ and $v_{i+k+1}$ do not see each other, we then
have $\gamma<\pi$. Thus, $v_{i},v_{j}$ and $v_{i+k+1}$ are the
vertices of a triangle in counter-clockwise order. The inner angles
of this triangle must sum up to $\pi$, i.e., $\measuredangle_{v_{i}}(v_{j},v_{i+k+1})+\measuredangle_{v_{j}}(v_{i+k+1},v_{i})+\measuredangle_{v_{i+k+1}}(v_{i},v_{j})=\pi$.
As $\gamma=\measuredangle_{v_{j}}(v_{i+k+1},v_{i})$, it is enough
to show that $\alpha<\measuredangle_{v_{i}}(v_{j},v_{i+k+1})$ and
$\beta<\measuredangle_{v_{i+k+1}}\!\left(v_{i},v_{j}\right)$ when
$v_{i}$ does not see $v_{i+k+1}$.

To see that this is the case, consider the right part of Figure~\ref{fig:angles--induction_step}.
The part of the boundary of $\mathcal{P}$ between $v_{i+k+1}$ and
$v_{i}$ (in counter-clockwise order) is the only part of the boundary
that can block the line-of-sight between $v_{i}$ and $v_{i+k+1}$.
This is because the line-of-sight from $v_{j}$ to $v_{i}$ and from
$v_{j}$ to $v_{i+k+1}$ must remain unblocked, and $\measuredangle_{v_{j}}(v_{i+k+1},v_{i})<\pi$.
But this means that there must be vertices from $\left\{ v_{i+k+2},v_{i+k+3},\ldots,v_{i-1}\right\} $
inside the triangle $v_{i},v_{j},v_{i+k+1}$. The $(r_{i}^{\left(k\right)}+1)$-th
edge of $v_{i}$ must lead to the ``right-most'' of these vertices.
Hence $\alpha<\measuredangle_{v_{i}}(v_{j},v_{i+k+1})$. Similarly,
it follows that $\beta<\measuredangle_{v_{i+k+1}}(v_{i},v_{j})$,
and thus $\alpha+\beta+\gamma<\pi$. This concludes the proof that
$\alpha+\beta+\gamma=\pi$ if and only if $v_{i}$ sees $v_{i+k+1}$.

As we derived a necessary and sufficient condition of when to add
an edge $\left\{ v_{i},v_{i+k+1}\right\} $ to the visibility graph,
the resulting algorithm computes the a solution. A naïve implementation
runs in time $O\!\left(n^{3}\right)$. We get the following theorem. 
\begin{thm}[\citep{DisserMihalakWidmayer/11}]
A simple polygon $\mathcal{P}$ is uniquely determined by its ordered
list of angle measurements. Moreover, there is a polynomial-time algorithm
that, given this data, reconstructs $G_{\mathrm{vis}}$ and hence
$\mathcal{P}$.\end{thm}
\begin{cor}
An angle agent can solve the visibility graph reconstruction problem,
even if restricted to moving along the boundary only.
\end{cor}

\subsection{Weaker agent models}

In the previous section we extended the basic agent model by a sensor
that allows to measure the angle between any any two edges incident
to the agent's location. We have shown that the angle agent can always
reconstruct $\mathcal{P}$ up to similarity (and thus $G_{\mathrm{vis}}$).
We actually weakened the model by restricting the agent to move along
the boundary only. As we are interested in understanding how much
information has to be gathered in order to reconstruct $\mathcal{P}$,
it is natural to ask whether our model is more powerful than actually
needed for this task. We will now briefly look at different ways of
weakening the angle agent further.

\begin{figure}
\centering{}\includegraphics{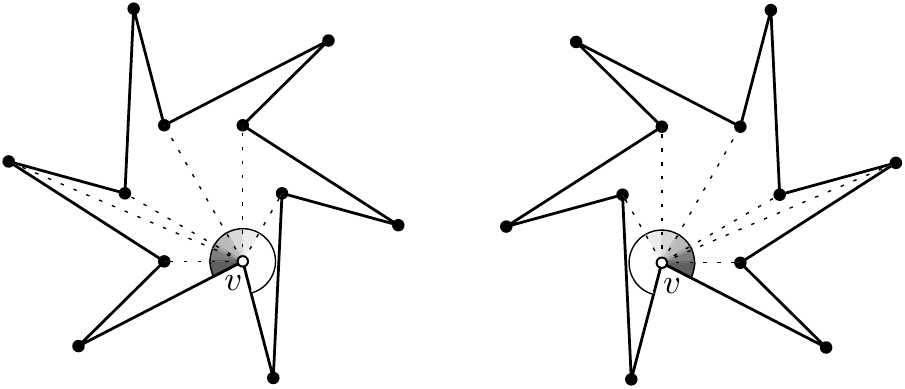}
\caption{Two polygons of different shape with the same set of angles at every
vertex. The two polygons are mirror images of each other.\label{fig:angles--order_of_angles_not_known}}
\end{figure}

\begin{figure}
\centering{}\includegraphics{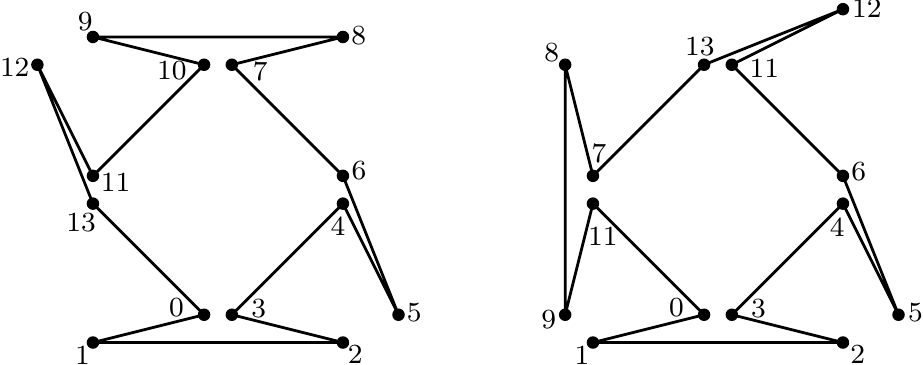}
\caption{Two different polygons with the same set of angle measurements. Every
angle measurement of a vertex in the left polygon also appears as
the angle measurement of a vertex in the right polygon. The order
in which these measurements occur along the boundary is different
however.\label{fig:angles--order_of_vertices_not_known}}
\end{figure}

A key feature of our model is the assumption that both the list of
measurements as well as the angles in each individual measurement
are given in counter-clockwise order. This comes naturally in the
context of our agent model. In terms of the geometrical formulation
of the problem however, we can ask whether we really need this additional
structure. Let us assume for a moment that each measurement is unordered,
i.e., instead of yielding a list of angles, each measurement yields
an unordered set of these same angles. Figure~\ref{fig:angles--order_of_angles_not_known}
gives an example of two different polygons with the same ordered list
of \emph{unordered angle measurements}. This means that if the measurements
are unordered, the polygon cannot always\emph{ }be reconstructed.
Similarly, Figure~\ref{fig:angles--order_of_vertices_not_known}
gives an example of two different polygons with the same \emph{unordered
set} of ordered angle measurements, i.e., the same ordered measurements
appear in both polygons, but not in the same order. This means that
without knowing the order in which the measurements occur along the
boundary, the polygon (or $G_{\mathrm{vis}}$) cannot be reconstructed
either.

Another feature of the model is that the angle agent is assumed to
know the total number of vertices $n$. It has been shown, however,
that the agent can even do without this prior knowledge of $n$~\citep{DisserMihalakWidmayer/10b}.
The difficulty in this case is that the agent cannot collect all information
in the beginning, as it would not know when it has completed its tour
of the boundary. The reconstruction can still be done by adapting
the algorithm we presented above. The main idea is to gradually identify
the vertices visible to $v_{0}$ in steps $k=1,2,\ldots$, by repeatedly
answering the question whether $v_{0}$ sees $v_{k}$. To answer this,
similarly to the algorithm above, the agent only needs to know the
subgraph of $G_{\text{vis}}$ induced by $\left\{ v_{0},\ldots,v_{k-1}\right\} $
and the subgraph of $G_{\text{vis}}$ induced by $\left\{ v_{1},\ldots,v_{k}\right\} $,
as well as the angle measurements at vertices $v_{0},v_{1},\ldots,v_{k}$.
This fact immediately yields an inductive approach, in which the agent
only needs to gather one new angle measurement in each step. After
all edges incident to $v_{0}$ have been identified, the global index
of the clockwise neighbor $v_{n-1}$ of $v_{0}$ has been obtained,
and thus $n$. Afterwards, the agent can either continue to use the
modified algorithm for all other vertices, or it can simply fall back
to the algorithm for the case where $n$ is known a priori.

Another way to reduce the amount of information available to the agent
is to weaken the angle measurement at each vertex. For example, the
agent might only be able to measure the inner angle of the polygon
at each vertex, i.e., the angle inside the polygon formed by the two
boundary edges adjacent to the vertex. It has been shown that in this
case the agent cannot always reconstruct $\mathcal{P}$~\citep{BiloDisserMihalakSuriVicariWidmayer/09}.
This remains true even if the agent can additionally decide which
vertices, among those currently visible, are neighbors on the boundary
(i.e., the agent can perceive the \emph{combinatorial visibility vector},
cf.~Section~\ref{sec:summary_mapping}).

\begin{figure}
\centering{}\includegraphics{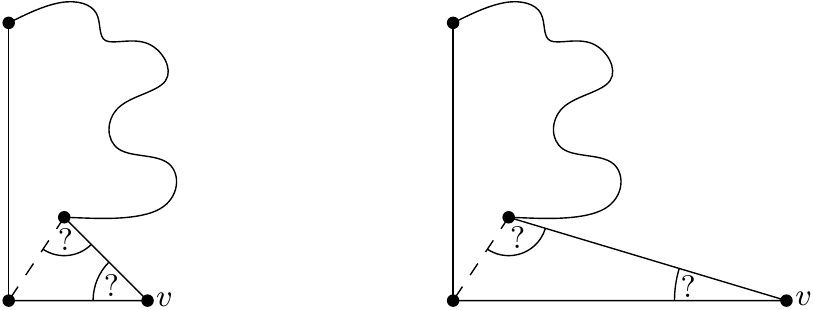} \caption{Two polygons with angle measurements that differ in exactly two angles.
This allows a single vertex $v$, which only sees its two neighbors
along the boundary, to be at different positions in both polygons.
Hence, the shape of a polygon cannot always be reconstructed if two
or more angles are unknown.\label{fig:angles--two_angles_not_known}}
\end{figure}

Finally, we could ask what if the agent can measure all angles except
for one angle (at any vertex). The missing angle can easily be computed
as the sum of all inner angles of $\mathcal{P}$ equals $(n-2)\pi$.
As soon as two angle values are missing, however, there can be more
than one polygon consistent with the angle measurements made by the
agent (cf.~Figure~\ref{fig:angles--two_angles_not_known}).

\section{Reconstruction by a freely moving agent\label{sec:base_graph}}

We approach the problem of reconstructing the visibility graph $G_{\mathrm{vis}}$
with an agent in the polygon $\mathcal{P}$ by conceptually splitting
it into a data collection phase and a computation phase. In the data
collection phase, the agent gathers all the information it can ever
extract from $\mathcal{P}$, i.e., the agent gathers information until
it does not need to move any further as it will not obtain new data
that way. The computation phase then consists of using the collected
information to derive $G_{\mathrm{vis}}$ computationally. Throughout
this section, if not mentioned otherwise, we will assume that the
agent is an extension of the basic agent model and knows the number
$n$ of vertices of $G_{\mathrm{vis}}$. We will refer to such an
agent as an \emph{agent with knowledge of $n$.}

\begin{figure}
\begin{centering}
\includegraphics[width=0.8\columnwidth]{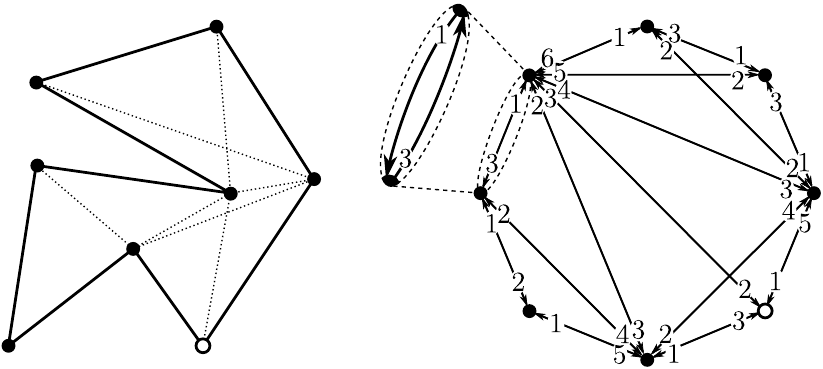} 
\par\end{centering}

\caption{A polygon with the corresponding directed and arc-labeled visibility
graph. Every bidirected edge represents two arcs of opposite orientation.\label{fig:basic labeling}}
\end{figure}

In the previous section we have introduced the angle agent for which
a single tour along the boundary provides enough information about
the polygon. In that case, the interesting part of the reconstruction
of $G_{\mathrm{vis}}$ is in the computation phase entirely. The situation
becomes more involved if a single tour around the boundary does not
suffice, and the agent thus needs to move across the polygon as well.
Recall that the basic agent model assumes the agent knows the counter-clockwise
order of all edges incident to its location, and thus has a way of
locally distinguishing them. This can be reflected in $G_{\mathrm{vis}}$
by replacing every undirected edge by two directed arcs in opposite
directions, labeling each outgoing arc of a node by its position in
the local counter-clockwise order (cf.~Figure~\ref{fig:basic labeling}).
The resulting labeling is a \emph{local orientation} in the sense
that every arc emanating from a node gets a label which is locally
unique. Without extra capabilities, this labeling is the only way
that an agent with knowledge of $n$ can distinguish arcs. In particular
the agent, in general, has no direct way of going back to where it
came from: after a move, it has no immediate way of distinguishing
the arc that leads back to its previous location. 

\begin{figure}
\begin{centering}
\includegraphics[width=0.5\columnwidth]{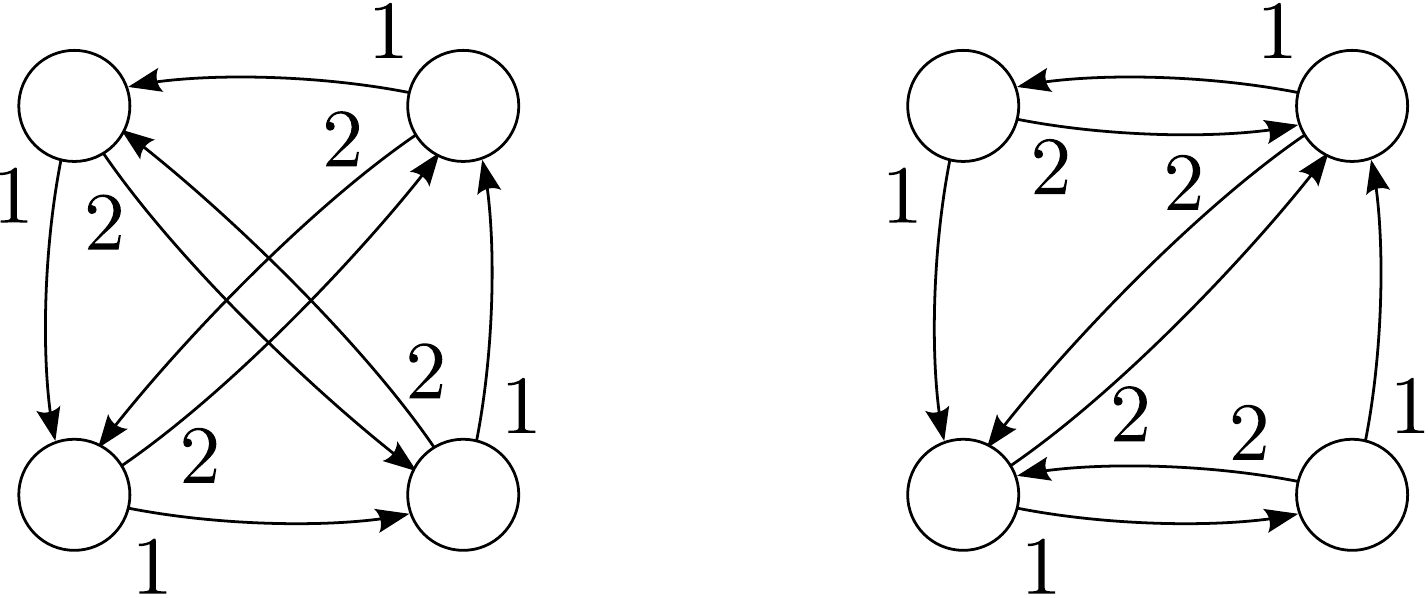}
\par\end{centering}

\caption{Two graphs that cannot be distinguished by an agent with knowledge
of $n$. \label{fig:Two-indistinguishable-graphs}}
\end{figure}

As the agent is moving along arcs of $G_{\mathrm{vis}}$, and as all
the data it can perceive is encoded in the labeling of $G_{\mathrm{vis}}$
(the knowledge of $n$ can be encoded easily), we can forget about
the underlying polygon $\mathcal{P}$ for a while. Instead, we view
our setting as an exploration of the directed, strongly connected,
and locally oriented graph $G_{\mathrm{vis}}$. Figure~\ref{fig:Two-indistinguishable-graphs}
shows that general locally oriented graphs cannot be reconstructed
by an agent with knowledge of $n$ (the sensing of the agent in this
setting yields the labels of the arcs emanating from its location):
The figure gives two different graphs which yield the same observations,
no matter how the agent decides to move in each step. While this shows
that the reconstruction of general directed graphs is impossible,
the situation might be different for (directed and labeled) visibility
graphs, as the agent might be able to exploit particular properties
of the graph and its labeling in this case. The question whether this
is possible has not been answered yet. In fact, the complete characterization
of the properties of visibility graphs is still an open problem, even
though it has received considerable attention in the past \citep{Ghosh/07,GhoshGoswami/09}.

The remainder of this section outlines a general method which an agent
with knowledge of $n$ and certain additional capabilities can employ
to reconstruct $G_{\mathrm{vis}}$. First, Section~\ref{sub:finding G^star}
describes how an agent exploring any locally-oriented arc-labeled
graph $G$ (think of a graph with colored edges) can systematically
gather all information it can ever encounter, i.e., how the agent
can perform the data collection phase in general graphs. Section~\ref{sub:there is a clique}
then goes back to the exploration of $G_{\mathrm{vis}}$ and establishes
the existence of certain structural properties in the collected data
which can be used as a basis for the reconstruction of $G_{\mathrm{vis}}$.
Finally, Section~\ref{sub:look-back} sketches an application of
the general method to a setting in which the agent is equipped with
a look-back sensor, which enables the agent to retrace its movements.

While the general method as described here is constructive, its purpose
is to prove that with certain additional capabilities the visibility
graph reconstruction problem can be solved, albeit not necessarily
in an efficient way. In fact, without additional considerations, the
resulting algorithms will generally be very inefficient as they iterate
over large sets of graphs.

\subsection{Data collection phase\label{sub:finding G^star}}

\begin{figure}
\begin{centering}
\includegraphics[width=1\columnwidth]{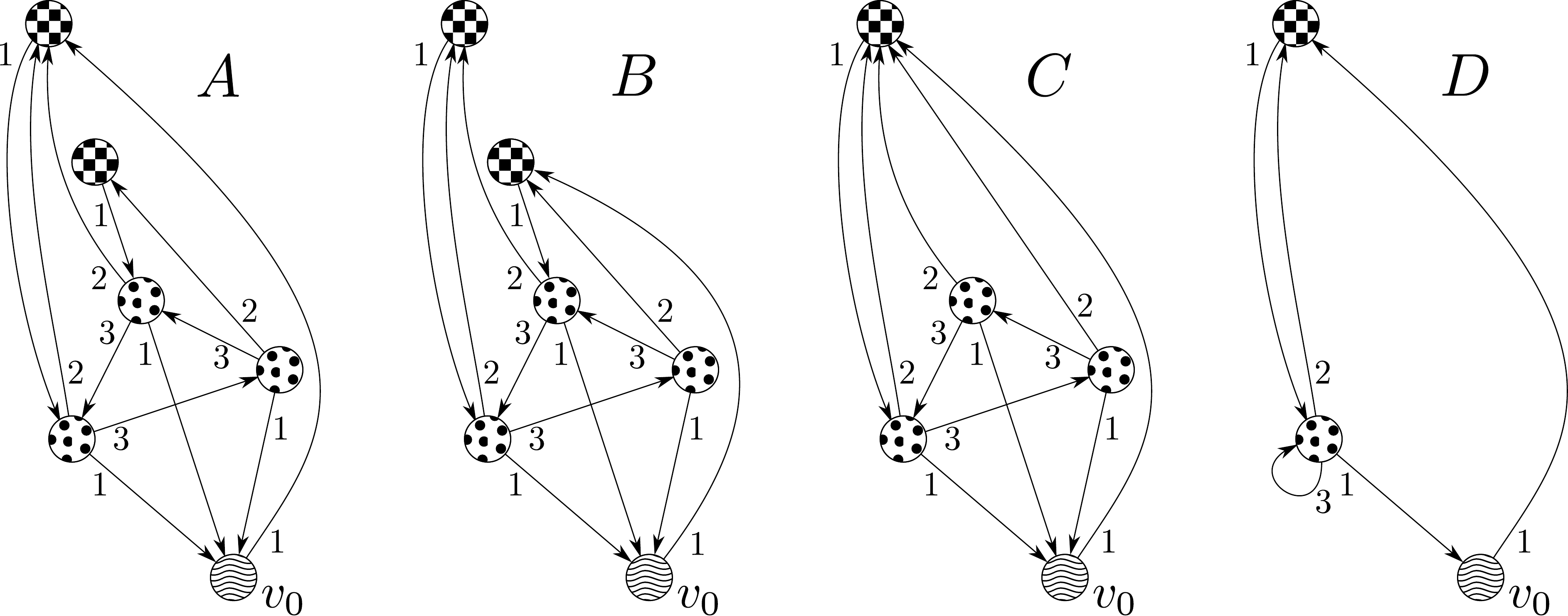} 
\par\end{centering}

\caption{Four graphs with local orientation that are indistinguishable to an
agent initially located at vertex $v_{0}$ in terms of observations
made by the agent along any walk. Vertices which appear indistinguishable
(depicted with individual patterns of the nodes) to the agent can
be merged in order to obtain smaller but still indistinguishable graphs.
\label{fig:ABCD indistinguishable}}
\end{figure}

As we saw in Figure~\ref{fig:Two-indistinguishable-graphs}, distinct
directed strongly connected arc-labeled (multi-) graphs can appear
indistinguishable to an agent moving along arcs and sensing arc-labels.
As another example consider an agent exploring graph $A$ from Figure~\ref{fig:ABCD indistinguishable}.
Inspecting the figure carefully, it should become clear that the same
sequences of labels appear along paths in each graph, and hence from
its observations alone, the agent has no way of knowing which of the
graphs it is exploring. As our agent has knowledge of the total number
of vertices $n$, it can exclude $C$ and $D$, but the agent is not
able to decide whether it is exploring $A$ or $B$. So, what information
\emph{can }the agent infer? Observe that the graph $D$ is the only
graph of size at most three which yields the same observations as
$A$, no matter how the agent decides to move in each step. If the
agent knew this smallest indistinguishable graph $D$, it would not
need to move at all anymore, as it could simulate all movements and
sensing in $D$ instead of executing both explicitly in the graph.
In other words, the agent cannot obtain more information about $A$
than the information encoded in $D$ (apart from the size $n$ of
$A$, which it knows already initially).

An important result from the field of graph fibrations implies that
the smallest graph indistinguishable in this way from some directed,
strongly connected, arc-labeled graph $G$ is always unique (up to
isomorphism) \citep{BoldiVigna/02}. This graph is called the \emph{minimum
base graph} of $G$, and we denote it by $G^{\star}$. As a consequence
we have that if an agent exploring $G$ can extract $G^{\star}$,
this completes the data collection phase of the reconstruction problem.

It turns out that an agent with knowledge of $n$ can\emph{ }extract
$G^{\star}$ in every (directed, strongly connected, arc-labeled)
graph $G$, given its number of vertices. The agent can employ the
following strategy that was introduced and proved formally in \citep{ChalopinDasDisserMihalakWidmayer/11}.
The goal of the agent is to determine both $G^{\star}$ as well as
the (unique) vertex $v^{\star}$ of $G^{\star}$ that appears indistinguishable
from the agent's starting location $v_{0}$ in $G$. The agent starts
with the set $\mathcal{C}$ of all candidate pairs $(G^{\prime},v^{\prime})$
consisting of a graph $G^{\prime}$ of size at most $n$ and a vertex
$v^{\prime}$ of $G^{\prime}$. Starting from this set, the agent
removes more and more candidate pairs, until only $\left(G^{\star},v^{\star}\right)$
remains. Initially, since $G^{\star}$ is the \emph{minimum} base
graph of $G$, it is safe to eliminate all graphs from $\mathcal{C}$
for which a smaller, indistinguishable graph exists. The idea now
is to repeatedly select two of the remaining pairs in $\mathcal{C}$
and eliminate one, until eventually only $\left(G^{\star},v^{\star}\right)$
remains. Doing that requires the agent to physically move in $G$,
keeping track of the sequence of labels $\sigma$ of the edges traversed
since the beginning of the execution. In order to conclude which one
of two pairs $\left(G_{1},v_{1}\right),\left(G_{2},v_{2}\right)\in\mathcal{C}$
can be discarded, the agent first determines, for $i\in\left\{ 1,2\right\} $,
the vertex $v_{i}^{\prime}$ that would be reached when exploring
graph $G_{i}$ starting at $v_{i}$ and following the label-sequence
$\sigma$. It can be show that there is a label-sequence $\delta$
that appears along a path starting at $v_{1}^{\prime}$ in $G_{1}$
but does not appear along any path in $G_{2}$ that starts at $v_{2}^{\prime}$,
or vice-versa. Without loss of generality, assume that $\delta$ appears
along a path starting at $v_{1}^{\prime}$ in $G_{1}$. The agent
now attempts to execute physical moves according to $\delta$. Either
this fails because at some point there is no edge with the correct
label at the agent's location, or the agent successfully traces $\delta$.
In the former case, the agent can conclude that $\left(G_{1},v_{1}\right)\neq\left(G^{\star},v^{\star}\right)$
and may thus eliminate $\left(G_{1},v_{1}\right)$ from $\mathcal{C}$.
In the latter case, the agent may, similarly, eliminate $\left(G_{2},v_{2}\right)$
from $\mathcal{C}$. After eliminating one of the pairs, the agent
updates $\sigma$ and removes all pairs $\left(G^{\prime},v^{\prime}\right)$
from $\mathcal{C}$ for which no walk in $G^{\prime}$ starting at
$v^{\prime}$ has edge-labels $\sigma$. Overall, in each step, $\left(G^{\star},v^{\star}\right)$
will remain in $\mathcal{C}$, but at least one other pair will be
removed. This implies that, while the strategy may be laborious, it
is guaranteed to terminate eventually. Note that the above can easily
be adapted if only an upper bound on $n$ is known.

So far we assumed $G$ to be a general directed, strongly connected,
and arc-labeled graph. This means that all of the above carries over
to the exploration of visibility graphs, provided that the data that
the agent is able to perceive can be encoded in an arc-labeling of
the directed visibility graph $G_{\mathrm{vis}}$. This is, for instance,
the case for an agent that is able to measure angles as introduced
in Section~\ref{sec:boundary_only}. If $\alpha_{i}$ is the counter-clockwise
angle between the arcs $i$ and $i+1$ at some node, this information
can for example be encoded by extending the label of the $i$-th arc
at the node from $i$ to $\left(i,\alpha_{i}\right)$. The more sophisticated
the sensors of the agent are, the more complex the arc-labels become,
and hence the minimum base graph $G_{\mathrm{vis}}^{\star}$ will
generally be larger. Intuitively, observing more local data results
in previously indistinguishable vertices becoming distinguishable.
For an agent with powerful sensors, this might mean that $G_{\mathrm{vis}}^{\star}=G_{\mathrm{vis}}$
and, consequently, the above strategy already yields the visibility
graph itself. On the other hand $G_{\mathrm{vis}}^{\star}\neq G_{\mathrm{vis}}$
means that the polygon has symmetries with respect to the agent's
perception, which make it impossible to derive $G_{\mathrm{vis}}$
without further exploiting the geometric meaning of the collected
data.

\subsection{Computation phase\label{sub:there is a clique}}

In the following, we go back to the setting of an agent with knowledge
of $n$ that is exploring $G_{\mathrm{vis}}$. We assume $G_{\mathrm{vis}}$
to be directed and labeled according to the data the agent can perceive,
as described before. Since an agent exploring any labeled graph can
always determine the minimum base graph if it knows (an upper bound
on) the size $n$ of $G_{\mathrm{vis}}$, the computation phase reduces
to: Given the minimum base graph $G_{\mathrm{vis}}^{\star}$ of $G_{\mathrm{vis}}$
and (a bound on) the size $n$ of $G_{\mathrm{vis}}$, compute $G_{\mathrm{vis}}$.

\begin{figure}
\begin{centering}
\includegraphics[width=1\columnwidth]{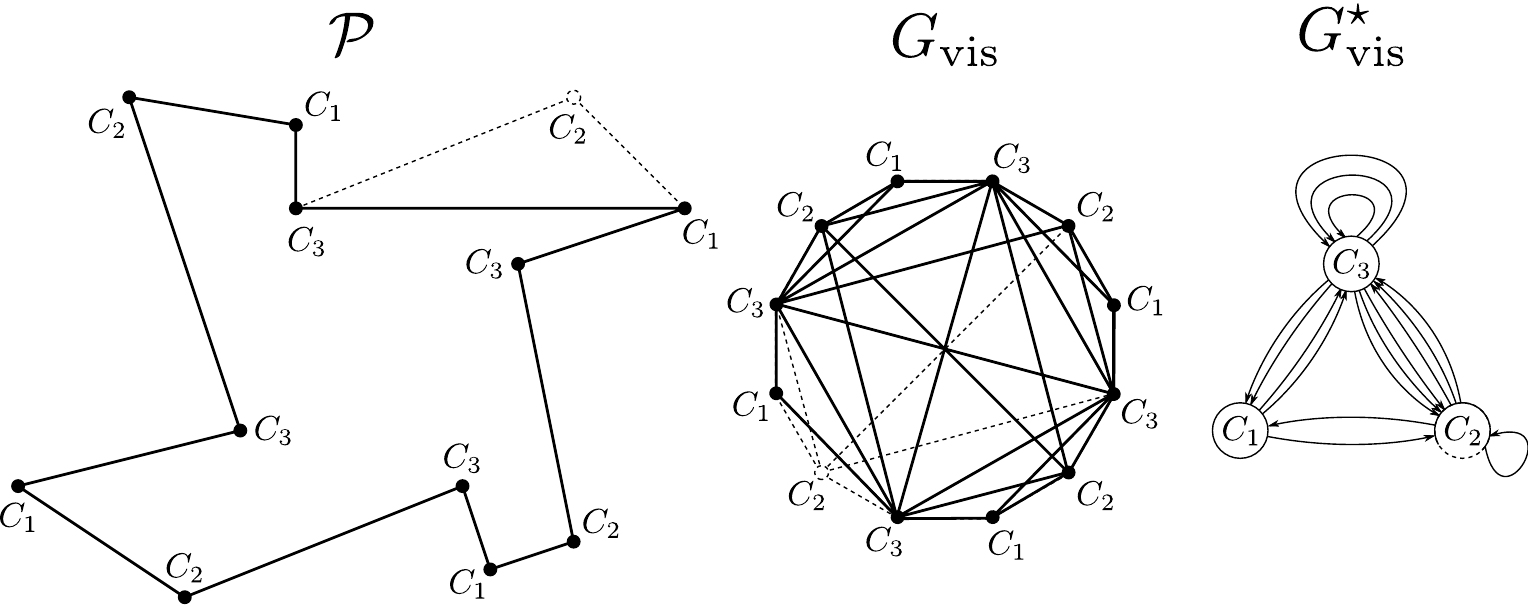} 
\par\end{centering}

\caption{A polygon $\mathcal{P}$ together with its visibility graph $G_{\mathrm{vis}}$
and the minimum base graph $G_{\mathrm{vis}}^{\star}$. Here $n=12$
and $k=3$, and every class contains $12/3=4$ vertices. Cutting off
a single vertex of $\mathcal{P}$ (dashed parts) cannot easily be
reflected in $G_{\mathrm{vis}}^{\star}$. In fact, in this example,
the new base graph would need to have eleven vertices. \label{fig:cutting ear}}
\end{figure}

Let us start by analyzing the relation between $G_{\mathrm{vis}}^{\star}$
and $G_{\mathrm{vis}}$. Observe that the vertices of $G_{\mathrm{vis}}^{\star}$
are pairwise distinguishable, i.e., the agent can distinguish which
of two vertices of $G_{\mathrm{vis}}^{\star}$ it is located at by
making a finite number of moves. This is because $G_{\mathrm{vis}}^{\star}$
is the smallest graph that is indistinguishable from $G_{\mathrm{vis}}$:
if two vertices of $G_{\mathrm{vis}}^{\star}$ would be indistinguishable,
a smaller graph could be obtained by merging the two vertices, i.e.,
by removing one of the vertices and connecting all of its incoming
arcs to the other vertex instead (cf. the transformation from $B$
to $C$ in Figure~\ref{fig:ABCD indistinguishable}). Therefore,
we naturally associate every vertex $v$ of $G_{\mathrm{vis}}$ to
the unique vertex of $G_{\mathrm{vis}}^{\star}$ which is indistinguishable
from $v$. With respect to indistinguishability, every vertex of $G_{\mathrm{vis}}^{\star}$
represents an equivalence class of vertices in $G_{\mathrm{vis}}$.
In terms of the agent's movements in $G_{\mathrm{vis}}$ this means
that starting on two different vertices of the same class with identical
movement decisions the agent will again end up in vertices of the
same class. This is evident, as the movements can be simulated in
$G_{\mathrm{vis}}^{\star}$ and there they result in the exact same
walk. Recall that the labeling of $G_{\mathrm{vis}}$ encodes the
position of each outgoing arc at a node in counter-clockwise order.
As the counter-clockwise ordering starts on the boundary of the polygon,
boundary arcs can be distinguished from non-boundary arcs in $G_{\mathrm{vis}}^{\star}$.
Both Hamiltonian cycles, one in clockwise and one in counter-clockwise
direction, induced by the boundary of $\mathcal{P}$ correspond to
(possibly smaller) Hamiltonian cycles in $G_{\mathrm{vis}}^{\star}$,
and thus the sequence of classes encountered along the boundary must
repeat $n/k$ times, where $k$ is the number of vertices of $G_{\mathrm{vis}}^{\star}$
(cf.~Figure~\ref{fig:cutting ear}). Consequently, all classes must
be of equal size.

\begin{figure}
\begin{centering}
\includegraphics[width=1\columnwidth]{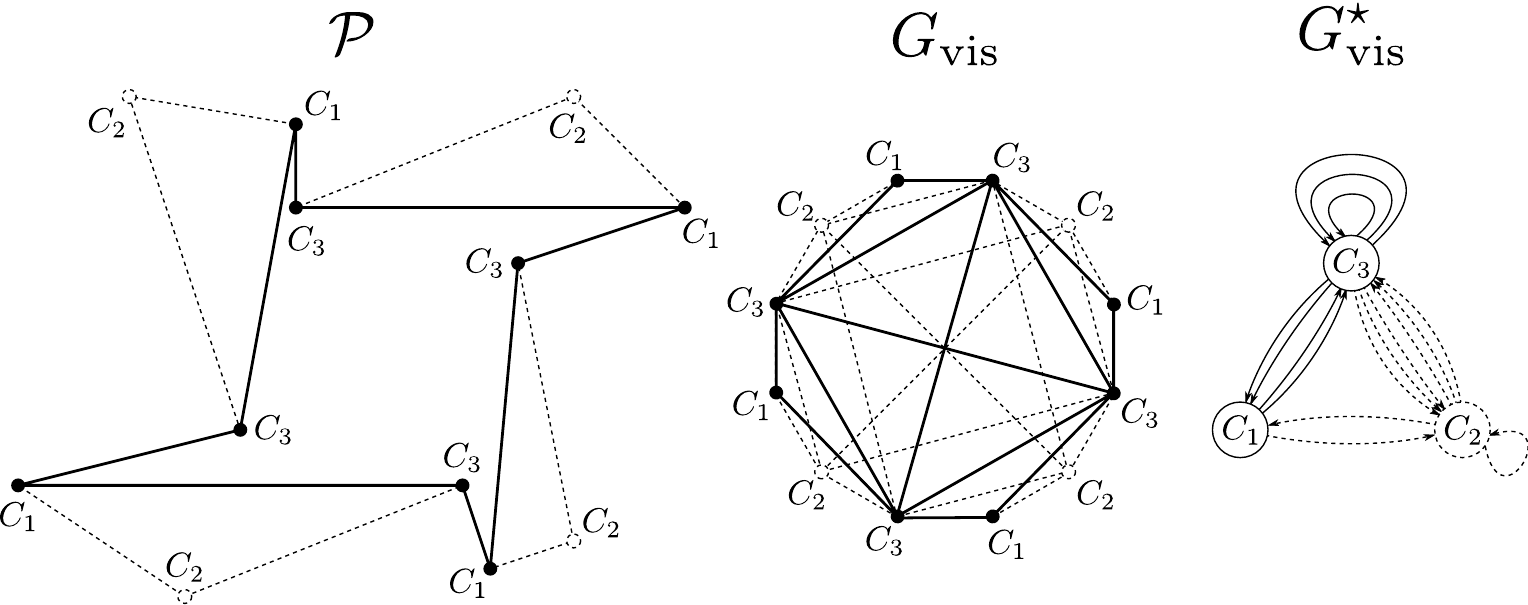} 
\par\end{centering}

\caption{Cutting off an entire class of ears can easily be done in $G_{\mathrm{vis}}^{\star}$
by removing the corresponding vertex entirely.\label{fig:Cutting a class}}
\end{figure}

Consider a node $v_{i}$ of $G_{\mathrm{vis}}$ such that $v_{i}$'s
neighbors on the boundary see each other, a so-called \emph{ear. }In
the underlying polygon, an ear is a vertex that does not obstruct
any lines of sight, i.e., an ear is a vertex that can be``cut off''
in order to obtain a smaller (but still simple) polygon (cf.~Figure~\ref{fig:cutting ear}).
As every polygon has at least one ear, this might suggest a recursive
approach for the reconstruction: cut off an ear, recurse, and glue
the ear back on. Recall however that we are operating on $G_{\mathrm{vis}}^{\star}$.
Even if we had a way of finding an ear, it is not clear how to cut
it off, i.e., how to obtain a graph indistinguishable to the visibility
graph of the new polygon from $G_{\mathrm{vis}}^{\star}$ (cf.~Figure~\ref{fig:cutting ear}).
Assume, however, that the following property holds. 
\begin{quote}
\textbf{Property ($\pentagon$): }For every ear $v_{i}$ of $G_{\mathrm{vis}}$,
all vertices in the same class as $v_{i}$ are ears. 
\end{quote}
Then, of course, it is still not clear how to cut off a single ear
in $G_{\mathrm{vis}}^{\star}$. Now, as all vertices in the same class
are ears as well, we can simply cut off all of them at once by removing
the corresponding vertex in $G_{\mathrm{vis}}^{\star}$ entirely.
The result is a graph which makes it possible to simulate an agent
moving inside the subpolygon obtained by cutting off the ears from
$\mathcal{P}$ (cf.~Figure~\ref{fig:Cutting a class}). As mentioned
above, every polygon has at least one ear. This means that, assuming
we could find these ears, we could repeatedly cut off entire classes
of ears, obtaining smaller and smaller polygons. At some point there
would be only one class $C^{\star}$ left in $G_{\mathrm{vis}}^{\star}$.
The corresponding polygon has to have at least one ear and by ($\pentagon$)
all its vertices must hence be ears. This can only be true for a convex
polygon. But that means that the vertices in $C^{\star}$ mutually
see each other, or in other words $C^{\star}$ forms a clique in $G_{\mathrm{vis}}$.
The result is the following theorem.

\begin{thm}[\citep{ChalopinDasDisserMihalakWidmayer/11,ChalopinDasDisserMihalakWidmayer/11c}]
If $\left(\pentagon\right)$ holds, then there is a class of vertices
that forms a clique in $G_{\mathrm{vis}}$.
\end{thm}
Assume that $\left(\pentagon\right)$ can be shown for an agent model.
Even if the agent has no way of finding ears, and thus no way of constructing
$C^{\star}$ as we did above, it can use the fact that there has to
be at least one class that forms a clique in $G_{\mathrm{vis}}$.
Such a class can immediately be found if $n$ is known, by inspecting
$G_{\mathrm{vis}}^{\star}$, because a class that forms a clique appears
as a vertex with $n/k-1$ self-loops in $G_{\mathrm{vis}}$ (cf.~Figure~\ref{fig:Cutting a class}).
Of course, no other vertex of $G_{\mathrm{vis}}^{\star}$ can have
more self-loops, and hence the total number of vertices $n$ is encoded
in $G_{\mathrm{vis}}^{\star}$. Because an upper bound on $n$ is
sufficient for finding $G_{\mathrm{vis}}^{\star}$, this means that
$n$ can always be inferred by an agent knowing only an upper bound
at the start.

There is no general way how to use the existence of a class that forms
a clique in the reconstruction of $G_{\mathrm{vis}}$. In some cases
its existence might make it easy to reconstruct $G_{\mathrm{vis}}$
directly. In other cases it might be possible to identify ears, and
therefore it might be possible to explicitly construct $C^{\star}$
by repeatedly cutting off classes of vertices. Starting with the clique
$C^{\star}$, the visibility graph could then possibly be built by
adding back the other classes one at a time.

\subsection{Example: agent with look-back sensor\label{sub:look-back}}

\begin{figure}
\begin{centering}
\includegraphics[width=0.8\columnwidth]{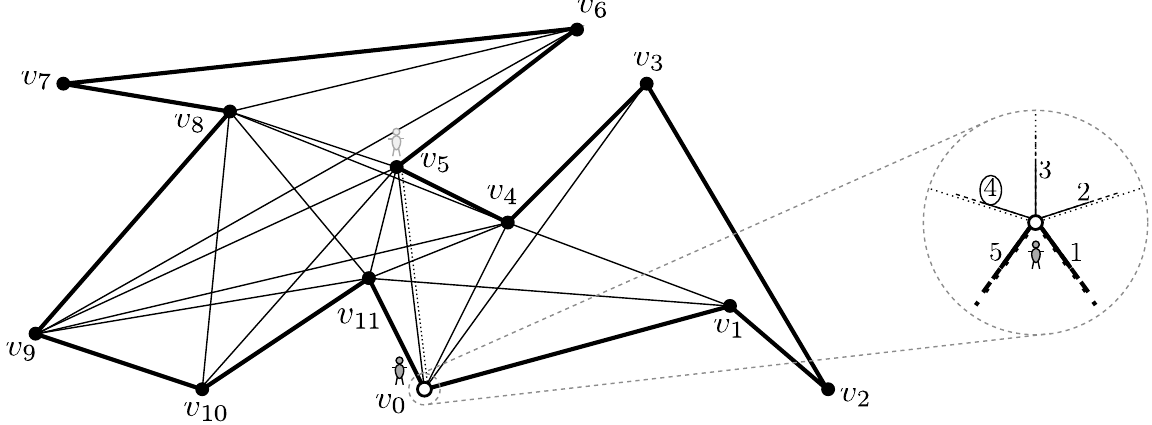} 
\par\end{centering}

\caption{The perception of the agent with look-back capability after its move
from $v_{5}$ to $v_{0}$. The agent can distinguish the arc leading
back to $v_{5}$, i.e., it knows its index ``4'' in the local ordering.\label{fig:look-back-perception}}
\end{figure}

A natural way for an agent to systematically collect information in
a graph-like environment is to traverse all possible walks in the
order of their length. However, without additional capabilities, an
agent with knowledge of $n$ encounters a major difficulty when trying
to perform a systematic search: after a move, the agent has no direct
way to return to where it came from, unless the move was along the
boundary. An way of avoiding this difficulty is to equip the agent
with an additional sensor which perceives the label of the arc that
leads back to the agent's previous location (cf.~Figure~\ref{fig:look-back-perception}).
We refer to an agent with this capability and knowledge of $n$ as
a \emph{look-back agent. }As there is no restriction on the memory
usage of even the basic agent, the look-back agent is able to backtrack
along walks of arbitrary length and thus to systematically traverse
all walks of a fixed length. But how much does this help for the reconstruction
of the visibility graph of a polygon? 

It turns out that, surprisingly, the capability of looking back alone
already empowers the agent to solve the reconstruction problem. The
general method that we have seen in the previous subsection is a generalization
of the proof strategy used in \citep{ChalopinDasDisserMihalakWidmayer/11c}
for a look-back agent. This section illustrates the general method
by applying it back to this original setting. While some parts of
the proof follow immediately from the general method, others have
to be proved individually for the particular setting. This section
aims to illustrate this interplay. In general, if the method is successful,
it yields a constructive proof that the visibility graph reconstruction
problem can be solved with a particular configuration of additional
capabilities of the agent. The method provides an algorithm that can
only guarantee an exponential running time. Making use of the particular
nature of the data available to an agent, however, the algorithm can
be sometimes improved to run in subexponential time (as discussed
above).

\begin{figure}
\begin{centering}
\includegraphics[width=0.8\columnwidth]{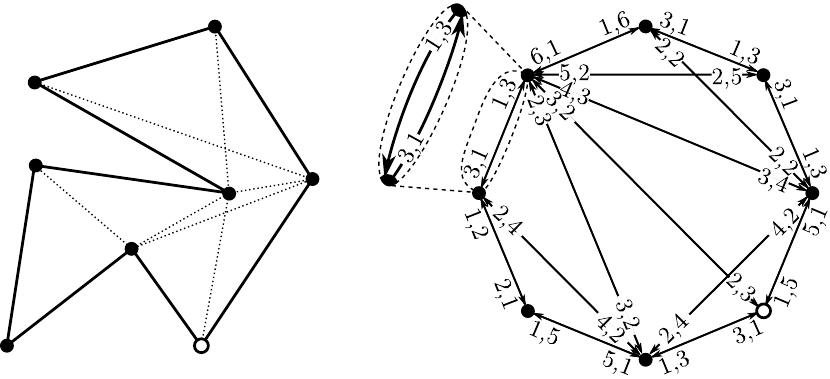} 
\par\end{centering}

\caption{An illustration of how the look-back capability can be encoded in
the arc-labeling of $G_{\mathrm{vis}}$ from Figure~\ref{fig:basic labeling}.
The highlighted vertex is an ear and has a path with labels $\left(1,5\right),\left(4,2\right)$
and a path with labels $\left(3,1\right),\left(2,4\right)$, where
3 and 5 are the degrees of the vertex itself and its counter-clockwise
neighbor, respectively.\label{fig:Encoding-the-look-back} }
\end{figure}

Applying the general method involves the following three steps that
are discussed in the following for the setting of a look-back agent: 
\begin{enumerate}
\item Encoding the sensing of the agent in an arc-labeling of $G_{\mathrm{vis}}$;
\item Proving Property $\left(\pentagon\right)$;
\item Using the existence of a class that forms a clique in $G_{\mathrm{vis}}$
for the reconstruction.
\end{enumerate}
For the first step, we need to extend the label of each arc in such
a way that the agent can identify which arc to use in order to back-track
a move. A straight-forward way of doing that is to take the standard
labeling for the basic agent (cf.~Figure~\ref{fig:basic labeling})
and extend the label of each arc $\left(u,v\right)$ by appending
the label of the arc $\left(v,u\right)$ as depicted in Figure~\ref{fig:Encoding-the-look-back}.
After moving along an edge with label $(a,b)$, a look-back agent
can back-track its move by moving along the\emph{ }arc with label
$(b,a)$. Recall that the original labeling was already a local orientation,
and thus the extended one is a local orientation as well.

The next step is to show Property $\left(\pentagon\right)$. One can
(easily) show that a node $v_{i}$ is an ear if and only if its counter-clockwise
neighbor along the boundary, $v_{i+1}$, has an edge with label $\left(d_{i+1}-1,2\right)$,
where $d_{i+1}$ is the degree of $v_{i+1}$. In other words, there
is a path of length two in $G_{\mathrm{vis}}$ which starts at $v_{i}$
and is labeled $\left(1,d_{i+1}\right),\left(d_{i+1}-1,2\right)$
(or equivalently, there is a path labeled $\left(d_{i},1\right),\left(2,d_{i+1}-1\right)$),
cf.~Figure~\ref{fig:Encoding-the-look-back}. On the other hand,
two nodes in the same class are indistinguishable, i.e., every sequence
of labels that occurs along a path starting at one of them also occurs
on a path starting at the other. In conclusion, if a node $v_{i}$
is an ear, it has a path labeled $\left(1,d_{i+1}\right),\left(d_{i+1}-1,2\right)$.
Hence all vertices in the class of $v_{i}$ have such paths. And hence
all these vertices are ears as well.

\begin{figure}
\begin{centering}
\includegraphics[width=0.6\columnwidth]{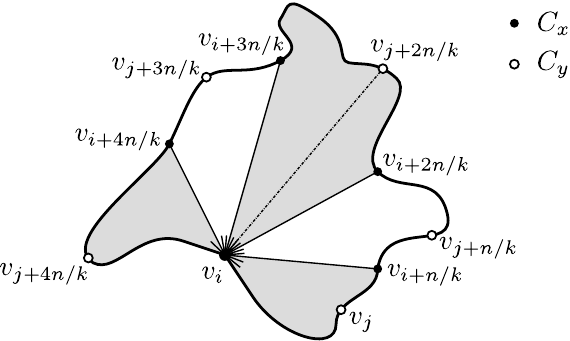} 
\par\end{centering}

\caption{A vertex $v_{i}$ in a class $C_{x}$ that forms a clique in $G_{\mathrm{vis}}$
has an arc to every other vertex of $C_{x}$. These arcs divide $\mathcal{P}$
into sectors (depicted as the alternating shaded and white areas),
such that each class appears exactly once in every sector. From $G_{\mathrm{vis}}^{\star}$,
the agent knows to which class $C_{y}$ an arc leads, and it also
knows the sector into which the arc falls. This is enough to uniquely
identify the exact target of the arc. Here, $k=5$ and $j=i+\left(\left(y-x\right)\mathrm{\, mod\,}k\right)$.\label{fig:sectors}}
\end{figure}

Let $C_{1},C_{2},\ldots,C_{k}$ be the different classes of $G_{\mathrm{vis}}$
in the order in which they appear along the boundary, where $k$ is
the number of vertices of $G_{\mathrm{vis}}^{\star}$. At this point,
by the general method, we know that there must be a class which forms
a clique in $G_{\mathrm{vis}}$. The look-back agent can easily find
such a class $C_{x}$ by counting the number of self-loops that every
vertex in $G_{\mathrm{vis}}^{\star}$ has -- recall that a class is
a clique if and only if its vertex in $G_{\mathrm{vis}}^{\star}$
has the maximum number of $n/k-1$ self-loops. Knowing $C_{x}$ can
help the agent in the reconstruction, as we will illustrate. Observe
that $G_{\mathrm{vis}}^{\star}$ encodes to which class every arc
leads; reconstructing $G_{\mathrm{vis}}$ means identifying, for every
arc, to which vertex of its class the arc leads. For a node $v_{i}\in C_{x}$,
this can be done quite easily. The agent knows that $v_{i}$ sees
all vertices in $C_{x}\backslash\left\{ v_{i}\right\} $, and that
these vertices are distributed evenly along the boundary, so the arcs
leading to $C_{x}$ must lead to the vertices $v_{i+n/k},v_{i+2n/k},v_{i+3n/k},\ldots$,
in that order. The arcs leading to $C_{x}$ partition the boundary
of $\mathcal{P}$ into sectors, where each sector contains exactly
one vertex of each class (cf.~Figure~\ref{fig:sectors}). As the
agent knows to which sector each arc at $v_{i}$ leads, and every
class occurs exactly once in this sector, it can immediately deduce
to which vertex an arc leads. The strategy becomes much more involved
if $v_{i}\notin C_{x}$. We omit the discussion of this case here
and refer to~\citep{ChalopinDasDisserMihalakWidmayer/11c} for more
details.

The general method gives an exponential time reconstuction algorithm
for a look-back agent. In this instance, the general approach can
be improved to yield a polynomial time reconstruction algorithm, as
was shown in \citep{ChalopinDasDisserMihalakWidmayer/11c}. A key
ingredient used in the proof is the fact that the information contained
in the tree of all walks starting from a node, the so-called \emph{view}
of the node, is already encoded in a subtree of depth $n-1$ \citep{Norris/95}.
Obtaining this finite subtree for every vertex is therefore enough
in order to group the vertices into equivalence classes with respect
to their view. As in the previous subsection, at least one of these
classes forms a clique in the visibility graph. Starting from this
clique, the entire visibility graph can be reconstructed. The straightforward
algorithm using this technique has exponential running time, because
it constructs all views up to depth $n$ in order to partition vertices
into equivalence classes. For this partitioning however, it is sufficient
to be able to distinguish between pairs of classes, i.e., between
every pair of distinct views. Thus, we can obtain a polynomial-time
algorithm that exploits this fact by constructing a walk for every
pair of classes, which it uses to distinguish between these two classes.

\begin{figure}
\begin{centering}
\includegraphics[width=0.8\columnwidth]{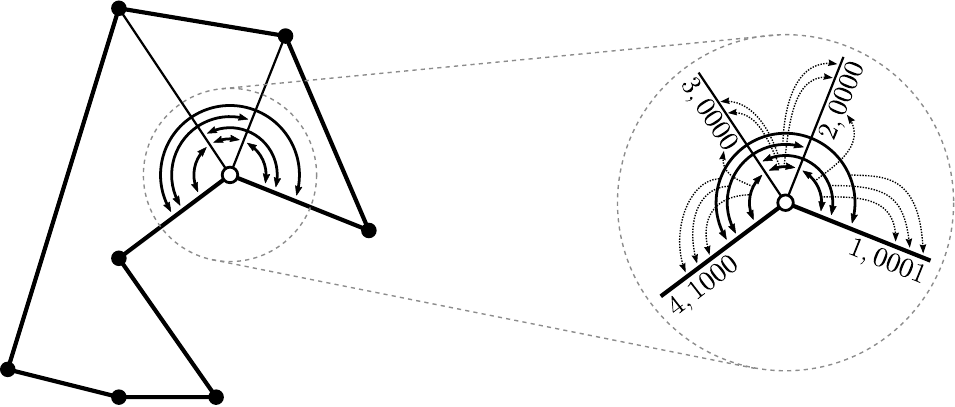} 
\par\end{centering}

\caption{Illustration of an angle-type sensor and how to encode it in the arc-labeling
of $G_{\mathrm{vis}}$.\label{fig:angle-types}}
\end{figure}

While instructive, it seems somewhat artificial to apply the general
reconstruction method to agents with look-back capability, since a
polynomial-time reconstruction algorithm for that setting exists.
The general method as described above was introduced in a different
context, in which the agent cannot look-back, but instead can roughly
measure angles. In that setting, for every pair of arcs emanating
from its location, the agent is able to distinguish whether the counter-clockwise
angle formed by these two arcs is convex ($\leq\pi$) or reflex ($>\pi$)
(cf.~Figure~\ref{fig:angle-types}). It is easy to see how to encode
this information in an arc-labeling: at every node $v_{i}$, every
outgoing arc is labeled with its position in counter-clockwise order
and, in addition, with a sequence $s\in\left\{ 0,1\right\} ^{d_{i}}$,
such that $s_{j}=0$ if and only if the arc forms a convex angle with
the $j$-th arc at the node (setting $s_{j}=0$ for the $j$-th arc).
More surprisingly, the other steps of the method can also be completed
\citep{ChalopinDasDisserMihalakWidmayer/11}, which shows that being
able to distinguish the ``type'' of the angle between arcs already
empowers an agent to reconstruct $G_{\mathrm{vis}}$. Without the
ability to backtrack movements, however, it seems doubtful that $G_{\mathrm{vis}}^{\star}$
can be constructed in polynomial time.

\section{Mapping polygons -- known results\label{sec:summary_mapping}}

This section is centered around Table~\ref{tab:known results}, which
gives a summary of the known results for the visibility graph reconstruction
problem. The underlying agent-models are modifications of the basic
agent model and differ in the type of sensors the agent is equipped
with additionally, the restrictions on the movements of the agent,
and the prior knowledge about the size of polygon. In the following,
the different components are explained briefly. For convenience, we
repeat the definitions of the sensors introduced in Sections \ref{sec:boundary_only}
and \ref{sec:base_graph}.
\begin{description}
\item [{angle~sensors}] The standard angle sensor (Section~\ref{sec:boundary_only})
measures all counter-clockwise angles between pairs of edges of $G_{\mathrm{vis}}$
which are incident to the agent's current location (cf.~Figure~\ref{fig:angle-sensor-kinds}~(a)).
The angle type sensor (Section~\ref{sub:look-back}) is the same
as the standard angle sensor, except that angles are not measured
exactly: for each angle, the angle type sensor only returns whether
this angle is convex ($\leq\pi$) or reflex ($>\pi$) (cf.~Figure~\ref{fig:angle-sensor-kinds}~(b)).
The inner angle sensor only measures the largest angle among all those
measured by the standard angle sensor (i.e., the angle formed by the
boundary edges of $\mathcal{P}$) (cf.~Figure~\ref{fig:angle-sensor-kinds}~(c)).
This angle, however, is returned exactly.
\begin{figure}
\begin{centering}
\includegraphics[width=0.8\columnwidth]{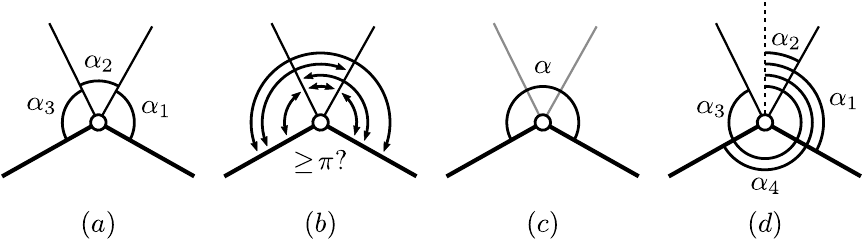} 
\par\end{centering}

\caption{Local perception with different kinds of angle sensors. From left
to right: standard angle sensor, angle type sensor, inner angle sensor,
angle sensor with compass.\label{fig:angle-sensor-kinds}}
\end{figure}

\item [{compass}] A compass provides a global reference direction that
is the same for every vertex. Also, the compass gives all the angles
between the global reference direction and each edge incident to the
current position (cf.~Figure~\ref{fig:angle-sensor-kinds}~(d)).
In particular, a compass yields all angle available with an angle
sensor.
\item [{cvv~sensor}] This sensor yields the \emph{combinatorial visibility
vector} of the agent's current location $v_{i}$: a binary vector
$\mathrm{cvv}\!\left(v_{i}\right)\in\left\{ 0,1\right\} ^{d_{i}+1}$
with the property that $\mathrm{cvv}_{0}\!\left(v_{i}\right)=\mathrm{cvv}_{d_{i}}\!\left(v_{i}\right)=1$,
and $\mathrm{cvv}_{j}\!\left(v_{i}\right)=1$ for $0<j<d_{i}$ if
and only if the $j$-th and $\left(j+1\right)$-th vertices that $v_{i}$
sees (in counter-clockwise order) are neighbors on the boundary (cf.~Figure~\ref{fig:cvv explained}).
Intuitively, the cvv sensor provides information between which of
the visible vertices (in counter-clockwise order) there are other
vertices which cannot be seen.
\begin{figure}
\begin{centering}
\includegraphics[width=0.6\columnwidth]{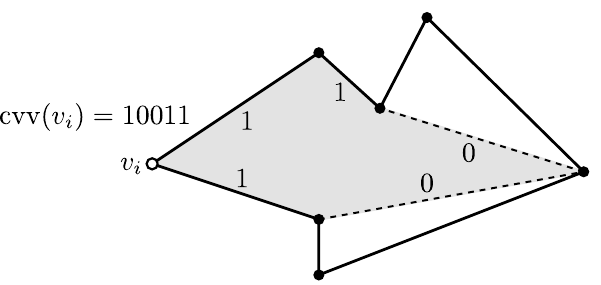} 
\par\end{centering}

\caption{The combinatorial visibility vector of a vertex $v_{i}$ encodes which
vertices visible to $v_{i}$ are neighbors along the boundary. The
shaded area is the subpolygon induced by the visible vertices of $v_{i}$.\label{fig:cvv explained}}
\end{figure}

\item [{distance~sensor}] The distance sensor simply measures the length
of every edge incident to the agent's current location, i.e., the
Euclidean distance from every visible vertex.
\item [{initial~info}] The basic model assumes that the agent has no knowledge
about the number of vertices of the polygon $n$. The model can be
altered further by assuming that $n$ is known, or that only an upper
bound $\bar{n}\geq n$ is known.
\item [{look~back}] If the previous move of the agent was from $v_{i}$
to $v_{j}$, the look-back sensor (Section~\ref{sub:look-back})
returns the index of the edge $\left(v_{j},v_{i}\right)$ in the ordered
list of edges that the agent perceives at $v_{j}$ (cf.~Figure~\ref{fig:look-back-perception}).
This enables the agent to undo past moves, as explained in Section~\ref{sub:look-back}.
\item [{movement}] We distinguish between two types of movement. One type
allows the agent to move freely along any edge of the visibility graph
which is incident to its current location. In the case of boundary
movement, the agent is restricted to moving along boundary edges only.
Section~\ref{sec:boundary_only} introduces the case of boundary
movement in more detail, while Section~\ref{sec:base_graph} studies
the case of free movement. 
\item [{pebble}] A pebble is an object used by the agent to mark vertices.
The pebble can be dropped by the agent at its current location and
can later be picked up again. The agent can distinguish the vertex
holding the pebble both when seen from afar as well as when the agent
reaches the vertex itself.
\end{description}
\begin{table}
\begin{centering}
\noun{Visibility Graph Reconstruction}
\par\end{centering}

\begin{centering}
\begin{tabular}{lcccccc}
\toprule 
 & initial &  &  & \multicolumn{3}{c}{results}\tabularnewline
sensors & info & movement &  & solvable & time & source\tabularnewline
\midrule
angles, compass & -- & boundary &  & yes & poly & \citep{BiloDisserMihalakSuriVicariWidmayer/09}\tabularnewline
angles & -- & boundary &  & yes & poly & \citep{DisserMihalakWidmayer/10b,DisserMihalakWidmayer/11}\tabularnewline
cvv, inner angles & $n$ & boundary &  & no &  & \citep{BiloDisserMihalakSuriVicariWidmayer/09}\tabularnewline
angle types &  & boundary &  & \emph{open} &  & \tabularnewline
distances &  & boundary &  & \emph{open} &  & \tabularnewline
pebble & -- & free &  & yes & poly & \citep{SuriVicariWidmayer/08}\tabularnewline
angle types, look-back & -- & free &  & yes & poly & \citep{BiloDisserMihalakSuriVicariWidmayer/09}\tabularnewline
look-back & $\bar{n}$ & free &  & yes & poly & \citep{ChalopinDasDisserMihalakWidmayer/11c}\tabularnewline
angle types & $\bar{n}$ & free &  & yes & exp & \citep{ChalopinDasDisserMihalakWidmayer/11}\tabularnewline
cvv, look-back & -- & free &  & no &  & \citep{BrunnerMihalakSuriVicariWidmayer/08}\tabularnewline
distances &  & free &  & \emph{open} &  & \tabularnewline
\emph{no sensors} & $n$ & free &  & \emph{open} &  & \tabularnewline
\bottomrule
\end{tabular}
\par\end{centering}

\caption{Summary of the cases in which an agent is known to be able/not able
to solve the visibility graph reconstruction problem. Note that a
polynomial running time in a setting where only an upper bound $\bar{n}\geq n$
is known a priori means polynomial in $\bar{n}$ rather than $n$.\label{tab:known results}}
\end{table}

\section{Related work\label{sec:related_work}}

In the previous sections we investigated \emph{agents} exploring \emph{polygons}
with the goal of \emph{reconstructing} a map of their environment.
We now extend our attention to related settings and point out some
prominent results. We in turn consider variations on the task to be
solved, the means by which data is acquired, and the nature of the
environment.

\subsection{Agents in polygons solving tasks other than mapping}

So far, we focused on the mapping problem, but mobile agents can be
used to perform a variety of other tasks in a polygon. One of the
first problems that was considered for mobile agents was the art-gallery
problem. We now exhibit this and a few other problems.

\paragraph{Art-gallery problem}

The \emph{art-gallery problem} asks to place guards at some vertices
of a simple polygon with $n$ vertices such that every point inside
the polygon is visible to at least one of the guards. The famous art-gallery
theorem asserts that every polygon can be guarded in this way by placing
at most $\lfloor n/3\rfloor$ guards~\citep{Chvatal/75,Fisk/78}.
While the theorem assumes full knowledge of the geometry of the polygon,
Ganguli et al.~\citep{GanguliCortesBullo/06} considered the art-gallery
problem on an initially unknown polygon, where the guards are autonomous
mobile agents. The guards were allowed to move freely inside the polygon
and to communicate over a distance as long as they see each other.
Ganguli et al.~showed that $\lfloor n/2\rfloor$ such mobile guards
can self-deploy at vertices such that the polygon is guarded. This
result raises the question whether the gap between $\lfloor n/2\rfloor$
and $\lfloor n/3\rfloor$ is inherent, due to the fact that the global
geometry of the polygon is initially unknown to the agents.

Suri et al.~\citep{SuriVicariWidmayer/08} showed that this is not
the case, and in fact $\lfloor n/3\rfloor$ guards can self-deploy
to guard the polygon. The authors considered the basic agent model
of Section~\ref{sec:model} and additionally equipped each agent
with a pebble (cf.~Section~\ref{sec:summary_mapping}) and the capability
to perceive the combinatorial visibility vector of a vertex (cf.~Section~\ref{sec:summary_mapping}).
We note that this agent model is weaker than the one that Ganguli
et al. employed. Suri et al.~showed that their guards can compute
the visibility graph of any simple polygon and thus also a triangulation
of the visibility graph. This allows to use the result of Fisk~\citep{Fisk/78},
who showed that covering a triangulation requires at most $\lfloor n/3\rfloor$
vertices, and hence a polygon can be guarded with $\lfloor n/3\rfloor$
guards if its triangulation is known. The proof is by first 3-coloring
the triangulation and then including the vertex of the least represented
color of each triangle into the covering.

\paragraph{Inferring the number of vertices of a polygon}

Any agent capable of reconstructing the visibility graph must implicitly
be able to infer the number of vertices $n$ of the polygon. Conversely,
if we can show that an agent cannot infer $n$, it follows that it
cannot reconstruct the visibility graph.

Naturally, computing $n$ is a trivial task for an agent with a pebble
-- an agent with a pebble can also infer the size of a polygon with
holes~\citep{SuriVicariWidmayer/08}. Similarly, the task is easy
when some vertex is locally distinguishable from all other vertices
and an upper bound on $n$ is known to the agent. The upper bound
is only needed in order for the agent to find such a vertex.

It has further been shown that without knowledge of $n$ or of an
upper bound on $n$, the agent can infer the size of a polygon in
the following three cases: (a) the agent is restricted to moving along
the boundary only and can perform angle measurements \citep{DisserMihalakWidmayer/10b},
(b) the agent can move freely along edges of the visibility graph,
can look-back and has an angle-type sensor \citep{BiloDisserMihalakSuriVicariWidmayer/09},
(c) the agent can move freely along edges of the visibility graph,
has an angle-type sensor and has a compass~\citep{BiloDisserMihalakSuriVicariWidmayer/09}.
An agent with combinatorial visibility and look-back capability, on
the other hand, cannot infer the size of a polygon in general~\citep{BrunnerMihalakSuriVicariWidmayer/08}. 

While the sensing model introduced in Section~\ref{sec:model} is
very simple and hence inherently robust, it requires that all visible
vertices can be perceived. Komuravelli and Mihalák~\citep{KomuravelliMihalak/09}
considered a faulty scenario in which it can happen that the agent
perceives two distant vertices as a single \emph{virtual} vertex (think
of vertices that appear very close to each other). They studied whether
an agent equipped with pebbles can infer the size of the polygon in
this setting. They showed that with a single pebble this is not possible,
conjectured that two pebbles are also still insufficient, and showed
that three pebbles allow computing the size of the polygon.

\paragraph{The rendezvous problem}

An important issue in distributed computing is how multiple autonomous
agents can cooperate with each other. In our context, a fundamental
question in whether $k>1$ agents moving inside the same polygon and
executing identical algorithms, can meet. Of course, the sensing of
the agents has to be extended such that agents can perceive each other.
A natural way of modeling this is to allow an agent to sense how many
agents are located at its current location and on each of the vertices
currently visible to the agent (this capability of the agent is sometimes
called \emph{strong} \emph{multiplicity detection}). All agents are
identical, which means that there is no way to distinguish them.

The \emph{strong rendezvous problem} requires the agents to gather
at an arbitrary vertex of the polygon after a finite number of steps.
Obviously, this is not always possible: Consider a convex polygon
with $n$ agents that are initially distributed over all $n$ vertices
of the polygon. There can be executions in which all agents act simultaneously,
and in which the distribution of the agents can therefore never change,
as every agent observes the same and thus all agents decide for the
same move at the same time.

The \emph{weak rendezvous problem} requires the agents to position
themselves such that all agents are mutually visible. In other words,
the locations of the agents have to form a clique in the visibility
graph. The weak rendezvous problem has been shown to be solvable for
agents that can look-back \citep{ChalopinDasDisserMihalakWidmayer/11c}
and for agents with angle-type sensor~\citep{ChalopinDasDisserMihalakWidmayer/11}.
The main idea for both results is to use the fact that there is a
class that forms a clique in the visibility graph. There is a unique
ordering among all those classes, and each agent can simply move to
a vertex of the first class of this ordering. We note that every agent
which can reconstruct the visibility graph can also solve the weak
rendezvous problem: Since the visibility graph is known, it is then
easy to simulate (on this graph) an agent with look-back capability
(for which the weak rendezvous problem has been shown to be solvable).
Notice that the agent that simulates the look-back agent only has
to physically move once it knows its final destination -- everything
else can be executed in memory. In particular, this implies that the
agents from Section~\ref{sec:boundary_only}, which move along the
boundary and measure angles, can solve the weak rendezvous problem.

\paragraph{Counting points inside a polygon}

Consider a simple polygon with $n$ vertices and a set of $k$ points
inside the polygon. A natural task for an agent in this context is
to count the number of points. Of course, we need to extend the sensing
model of the agent to make it aware of the points first: At any vertex
the agent now perceives the list of visible points and vertices ordered
in counter-clockwise order, rather than visible vertices only. The
agent is assumed to be able to distinguish between vertices and points.

Gfeller et al.~\citep{GfellerMihalakSuriVicariWidmayer/07} showed
that an agent with a pebble cannot approximate the number of points
within a factor of $2-\varepsilon$, for any $\varepsilon>0$. In
the following, a $\rho$-approximation of the number of points $k$
is an upper bound $z$ with $k\leq z\leq\rho k$. The results of Gfeller
et al.~imply that an agent knowing the \emph{vertex-edge visibility
graph} of the polygon (or being able to compute it), together with
its initial position in it, can compute a $2$-approximation of the
number of points. The vertex-edge visibility graph is a bipartite
graph with a node for every boundary edge and a node for every vertex
of the polygon. There is an edge between a node corresponding to a
boundary edge and a node corresponding to a vertex if at least part
of the boundary edge is visible to the vertex. We note that the vertex-edge
visibility graph induces the visibility graph \citep{ORourkeStreinu/98}.

It can be shown that the vertex-edge visibility graph is not needed
to compute a 2-approximation -- knowing the visibility graph instead
is already sufficient, at the cost of an exponential running time.
The idea is simply to iterate over all vertex-edge visibility graphs
that are compatible with the given visibility graph and run the 2-approximation
of Gfeller et al.~on each of those. The output is the smallest estimation
of the number of points encountered in the process.

\subsection{Reconstruction not involving agents}

In Section~\ref{sec:boundary_only}, we related the visibility graph
reconstruction problem for an angle agent inside a polygon to a general
problem from geometry which asks to reconstruct a simple polygon from
certain measurement data. In our case, we showed that a simple polygon
can be reconstructed, given its ordered list of angle measurements.
The general problem of reconstructing geometrical objects from various
kinds of measurement data has been studied in the past. We give a
brief overview of the most prominent results concerning the reconstruction
of polygons.

\paragraph{Constructing a consistent polygon}

There are two main variants of the general problem of deriving a polygon
from measurement data. The first variant asks to construct \emph{some}
polygon $\mathcal{P}^{\star}$ that is \emph{consistent} with the
data measured in the original polygon $\mathcal{P}$ (naturally, $\mathcal{P}$
is not part of the input). Being consistent means that this data could
have originated from a series of measurements in $\mathcal{P}^{\star}$
rather than in $\mathcal{P}$. Studies that consider this variant
of the reconstruction problem usually focus on the complexity of the
problem rather than its feasibility.

The first problem that was studied in this context asked to construct
a polygon which is consistent with a given visibility graph. This
problem is only known to be in PSPACE -- its complexity is still open~\citep{Everett/90}.
A related problem is the characterization of visibility graphs, which
as well is a long-standing open problem~\citep{Ghosh/07}.

\begin{figure}
\begin{centering}
\hfill{}\includegraphics[width=0.25\textwidth]{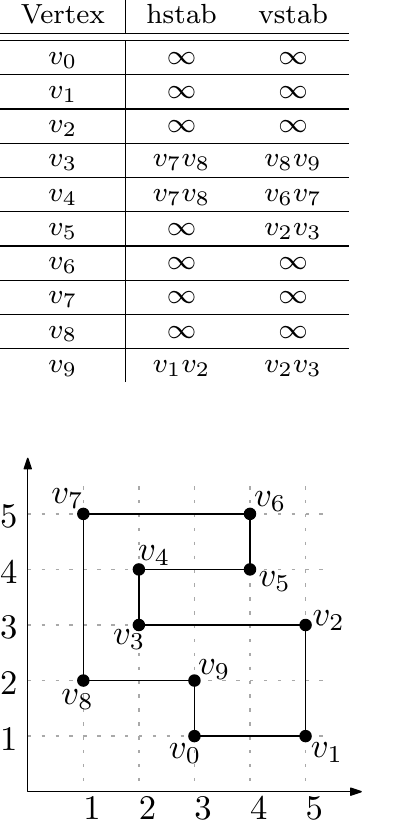}\hfill{}\includegraphics[width=0.3\columnwidth]{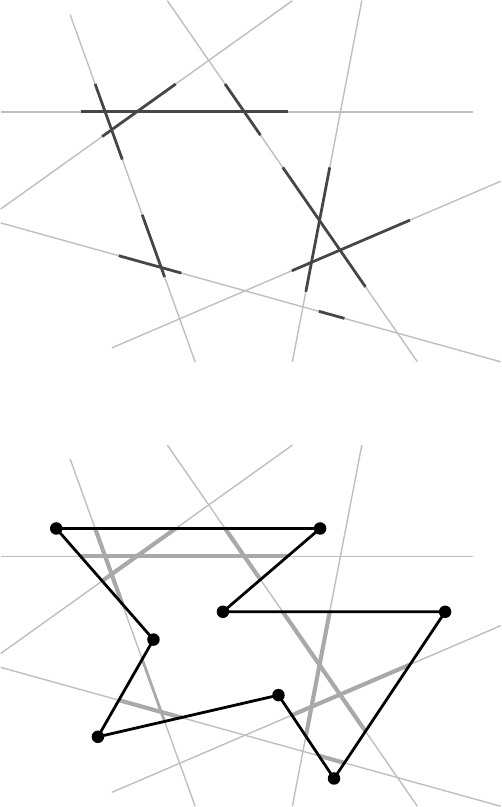}\hfill{}\includegraphics[width=0.3\columnwidth]{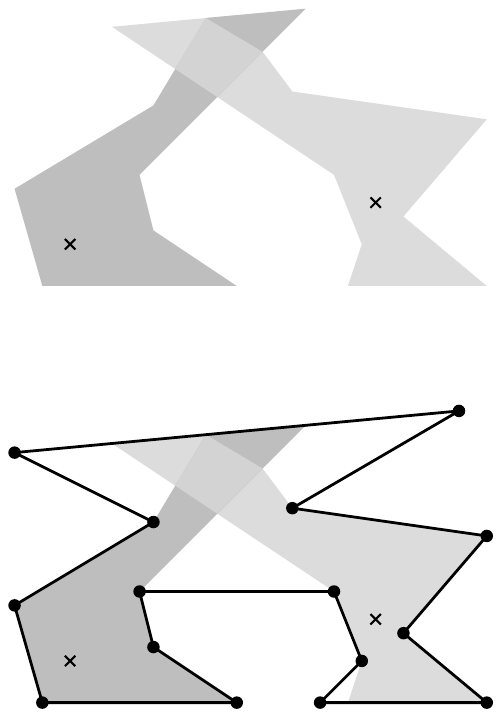}\hfill{}
\par\end{centering}

\caption{Constructing consistent polygons from different kinds of data. From
left to right: construction of orthogonal simple polygons from stabbing
information; construction of polygons from line intersections; construction
of simple polygons from visibility polygons of certain points in the
interior.\label{fig:Constructing-consistent-polygons-1}}
\end{figure}

Jackson and Wismath~\citep{JacksonWismath/02} studied the problem
of constructing an orthogonal simple polygon from the ``stabbing
information'' at all vertices (cf.~Figure~\ref{fig:Constructing-consistent-polygons-1}~(left)).
An orthogonal polygon is a polygon $\mathcal{P}$ for which every
boundary edge is either horizontal or vertical. The to-be-found orthogonal
polygon is assumed to have no three vertices lying on a vertical or
horizontal line, i.e., in particular, every vertex has a horizontal
and a vertical boundary edge adjacent to it. A \emph{horizontal stab}
of vertex $v$ is the horizontal ray starting at $v$ and going to
the ``opposite'' side of $v$'s horizontal boundary edge. A \emph{vertical
stab} is defined analogously. The \emph{stabbing information} of vertex
$v$ is the pair of boundary edges of $\mathcal{P}$ with which the
horizontal and the vertical stab of $v$ intersect first. If there
is no intersection, a placeholder ``phantom'' line-segment is provided
instead (denoted as $\infty$ in the figure). Jackson and Wismath
presented an algorithm with a running time of $O(n\log n)$ that computes
an orthogonal simple polygon $\mathcal{P}^{\prime}$ of size $n$
that is consistent with the given stabbing information at every vertex.
We note that there can be more than one simple polygon consistent
with this stabbing information.

Sidlesky et al.~\citep{SidleskyBarequetGotsman/06} considered a
similar construction problem in which all intersections of $\mathcal{P}$
with a given set of lines $\mathcal{L}$ are given (cf.~Figure~\ref{fig:Constructing-consistent-polygons-1}~(middle)).
One may assume that every boundary edge of $\mathcal{P}$ is intersected
by at least two lines from $\mathcal{L}$, as otherwise there are
infinitely many polygons $\mathcal{P}'$ that intersect $\mathcal{L}$
in the same way that $\mathcal{P}$ does. The authors present an exponential-time
algorithm that constructs all polygons $\mathcal{P}'$ that are consistent
with the given intersections, including non-simple polygons.

Biedl et al.~\citep{BiedlDurocherSnoeyink/09} considered various
types of measurements in a simple polygon, and considered the complexity
of the problem to decide whether or not there is a simple polygon
that is consistent with the given data. Examples for the measurements
considered are (1) a set of points on the boundary of the original
simple polygon, such that every boundary edge contains at least one
point, and (2) a set of visibility polygons, i.e., the regions of
the polygon that are visible from certain points in the polygon (for
the latter, cf.~Figure~\ref{fig:Constructing-consistent-polygons-1}~(right)).
If no restriction on the shape of the polygon is enforced, the problem
was shown to be NP-hard for each type of data considered in the study.
Polynomial-time algorithms were given for special cases only, such
as when the polygon is required to be orthogonal and monotone, or
star-shaped. Other special cases remain NP-hard, such as when the
polygon is required to be orthogonal but not monotone. 

\begin{figure}
\begin{centering}
\hfill{}\includegraphics[width=0.3\columnwidth]{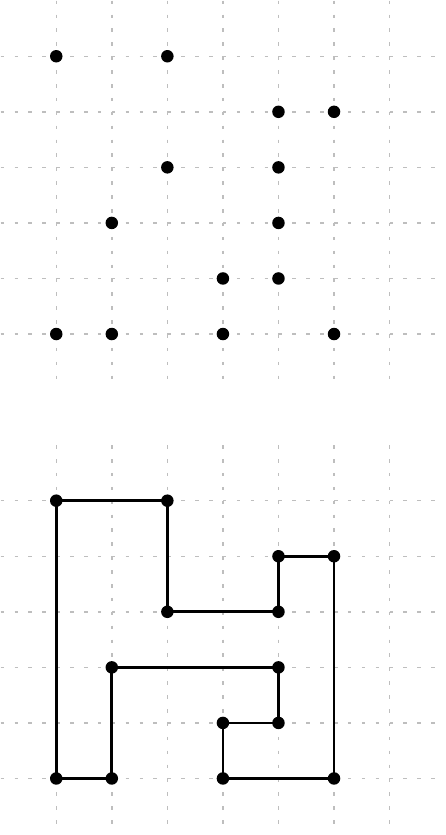}\hfill{}\includegraphics[width=0.3\columnwidth]{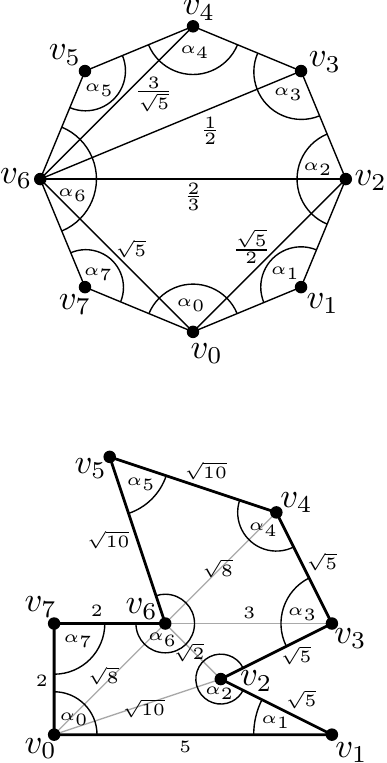}\hfill{}
\par\end{centering}

\caption{Reconstructing a polygon from different kinds of data. Left: reconstruction
of an orthogonal simple polygon from the coordinates of its vertices.
Right: reconstruction of a simple polygon from the triangulation of
its visibility graph with cross-ratios and inner angles. The top figure
shows the triangulation of the visibility graph, where each edge is
labeled by the cross-ratio of the corresponding quadrilateral. The
bottom picture shows the underlying triangulation of the polygon,
where each edge is labeled by its length.\label{fig:Reconstructing-uniquely-1}}
\end{figure}

Rappaport~\citep{Rappaport/86} considered the \emph{orthogonal-connect-the-dots}
problem: Given a set of points in the plane, find an orthogonal simple
polygon whose vertices coincide with these points (cf.~Figure~\ref{fig:Reconstructing-uniquely-1}~(left)).
Note that here the points are not given in a particular order. Rappaport
was able to show that the problem is NP-complete, when 3 consecutive
points on the boundary can lie on a horizontal or vertical line.

\paragraph{Reconstructing the polygon}

The second variant of the construction problem asks to reconstruct
\emph{the} \emph{original} polygon $\mathcal{P}$ from data measured
in $\mathcal{P}$. Solving this problem not only involves constructing
a consistent polygon, but also requires to show that among all polygons,
$\mathcal{P}$ itself is the only polygon that is consistent with
the data measured in $\mathcal{P}$. This type of reconstruction problem
arises naturally in the context of autonomous agents mapping polygonal
environments: It is not enough for an agent to construct \emph{some}
polygon which is consistent with its observations, the agent wants
to find the exact polygon it is located in. Here the focus is on the
question whether reconstruction is at all possible from the given
data, finding efficient algorithms is only of secondary interest.

Recall the orthogonal-connect-the-dots problem mentioned above (cf.~Figure~\ref{fig:Reconstructing-uniquely-1}~(left)).
O'Rourke~\citep{ORourke/88} considered the problem further and showed
that if no three consecutive vertices on the boundary of the polygon
lie on a vertical or horizontal line, then the coordinates of the
vertices determine the polygon uniquely. What is more, he showed that
in this case the problem is not NP-hard anymore, by providing an algorithm
that solves the reconstruction problem in time $O(n\log n)$.

In Section~\ref{sec:angles-solution} we focused on reconstructing
a simple polygon $\mathcal{P}$ from its ordered list of angle measurements,
and we showed that $\mathcal{P}$ is indeed the only polygon consistent
with this data. Moreover we developed a polynomial-time algorithm
that finds the original polygon $\mathcal{P}$. These results appeared
in \citep{DisserMihalakWidmayer/11}.

Coullard and Lubiw~\citep{CoullardLubiw/91} studied the problem
of deciding whether a given edge-weighted graph is the \emph{distance
visibility graph} of a polygon, i.e., a visibility graph with edge-weights
equal to the length of the corresponding line segments in the plane.
Note that this question is easy to decide in exponential time, by
trying out all Hamiltonian cycles and repeatedly triangulating the
graph based on the current cycle. If one of these triangulations can
be embedded in the plane as a simple polygon (and the other edge-weights
not used in the triangulation are consistent with the embedding),
then, and then only, the given graph is a distance visibility graph.
Furthermore, the resulting polygon is uniquely defined by the distances.
Coullard and Lubiw gave a necessary condition for a graph (without
distances) to be the visibility graph of a polygon, and, based on
this property, the authors proposed a polynomial-time algorithm that
decides whether a given edge-weighted graph is the distance visibility
graph of a simple polygon.

Snoeyink~\citep{Snoeyink/99} showed that every simple polygon $\mathcal{P}$
on $n$ vertices is uniquely determined by its triangulation given
as a graph, its \emph{inner angles}, and its $\left(n-3\right)$ \emph{cross-ratios}
(cf.~Figure~\ref{fig:Reconstructing-uniquely-1}~(right)). A \emph{cross-ratio}
of a non-boundary edge of a triangulation is defined in terms of quadrilateral\emph{s,
}i.e., pairs of triangles with a common edge in the triangulation.
If $a,b,c$ and $a,d,e$ are the edges of both its triangles in counter-clockwise
order, the \emph{cross-ratio} of a quadrilateral is the product of
the lengths of $b$ and $d$ divided by the product of the lengths
of $c$ and $e$. An \emph{inner angle} of a polygon at vertex $v$
is the angle inside $\mathcal{P}$ which is enclosed by the two boundary
edges adjacent to~$v$.

\subsection{Agents exploring other environments}

The basic agent as defined in Section~\ref{sec:model} moves along
the edges of the visibility graph of a simple polygon. We can readily
generalize this model for an agent that moves inside arbitrary graph-like
environments. In fact, some of the techniques that we introduced in
the previous sections were originally developed for this more general
setting. In the following, we mention some key results for the exploration
of graph-like environments. Research has mostly focused on determining
under which circumstances an agent can explore and reconstruct an
arbitrary connected graph.

Intuitively, exploring a general graph-like environment is more difficult
for the agent, since it cannot exploit the structure of the underlying
geometry. For example, the visibility graph of a simple polygon always
contains a Hamiltonian cycle (the boundary), and we have assumed in
this paper that the agent is able to distinguish the edges of this
cycle from other edges by their labels. In the general setting, there
is no such assumption on the labeling, except that the edges incident
to each vertex are assumed to be mutually distinguishable, i.e., the
graph is assumed to be locally oriented. On the other hand, in the
general setting, the agent is usually assumed to be able to look-back.
We will see shortly that in contrast to the exploration of visibility
graphs, in general graphs the ability to look-back alone does not
empower the agent to reconstruct the graph. Because of the look-back
ability, we can model the general setting as the exploration of an
undirected, edge-labeled graph of $n$ vertices and $m$ edges.

The exploration becomes straightforward when the nodes of the graph
are labeled by distinct identifiers. In this case, the map construction
problem is equivalent to the traversal problem, since the map can
easily be constructed as soon as the identifiers of the endpoints
of every edge are known. As the agent is capable of looking back,
it can retrace its movements and solve the traversal problem by employing
a conventional depth-first search algorithm. In fact, the depth-first
traversal is asymptotically optimal in terms of the number of required
moves, as it needs $\Theta(m)$ moves -- obviously no other solution
can do better as it has to travel along each edge at least once. A
modified version of the algorithm which takes $m+O(n)$ moves has
been proposed by Panaite and Pelc~\citep{PanPe99}. Another variation
of the traversal problem, the so-called \emph{piece-meal exploration},
has been studied by Awerbuch et al.~\citep{AwBS99}. Here the agent
can only execute a certain number of moves before it has to return
to its home-base (e.g. for refueling).

The map construction problem becomes more challenging in an anonymous
graph in which nodes are unlabeled. In this case, traversing the graph
is not equivalent to map construction except for some special classes
of graphs such as trees. If an upper bound on the diameter of the
graph (e.g.~$n$ or the diameter itself) is known, traversal can
always be achieved by trying all possible walks up to a length equal
to this bound. Note that the knowledge of an upper bound on the diameter
is necessary for terminating the traversal, without this bound the
agent can still travel all edges within a finite time, but it does
not know when this point is reached and can thus not stop after any
finite time. While we generally do not assume a bound on the memory
of the agent, it was shown that it needs at least $\Omega(\log{n})$
bits to traverse a graph of size~$n$~\citep{FraIPPP04}. The long
standing open question about the space complexity of graph traversal
was closed by Reingold~\citep{Reingold/05}, who showed a matching
upper bound of $O\!\left(\mathrm{log}\, n\right)$ bits on the required
memory.

\begin{figure}[htb]
\centering{}\includegraphics{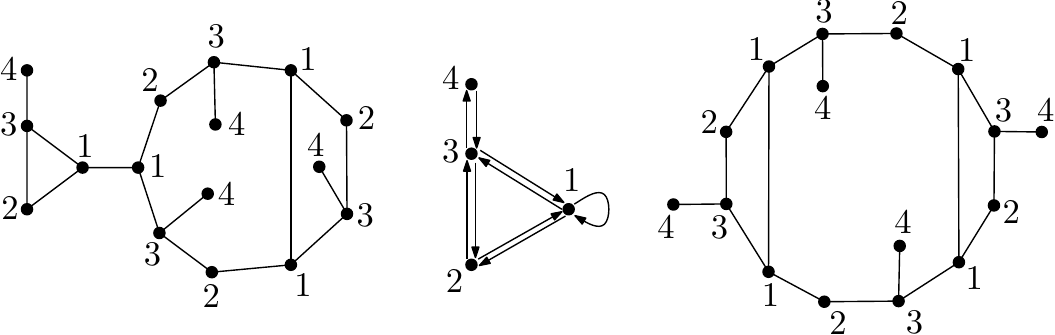} \caption{Two non-isomorphic graphs $G,H$ that are indistinguishable together
with their minimum base graph $B$. The numbers on the nodes represent
the equivalence classes with respect to symmetry. Edge labels are
omitted to not clutter the figure.\label{fig:nonrecognizable-1-1}}
\end{figure}

Even though all (connected) graphs can be systematically traversed
by an agent, the map construction problem cannot be solved in general.
We have seen examples for indistinguishable non-isomorphic directed
graphs in Section~\ref{sec:base_graph} (cf.~Figure~\ref{fig:ABCD indistinguishable}).
It turned out that in polygonal environments, the capability of ``looking
back'' empowers the agent to distinguish any pair of visibility graphs.
Intuitively, this does not carry over to general graphs due to symmetries
that can occur. As an example for indistinguishable undirected graphs,
consider the pair of graphs $G,H$ shown in Figure~\ref{fig:nonrecognizable-1-1}.
The two graphs are non-isomorphic, but an agent traversing graph $G$
makes the exact same observations as an agent traversing graph $H$,
provided that both agents start on corresponding vertices. Thus, both
these graphs are non-recognizable, i.e., the map construction problem
cannot be solved for either of them. The class of all recognizable
graphs has been characterized by Yamashita and Kameda~\citep{YamashitaKameda/96}.

A main problem to tackle when constructing a map is how to deal with
the observed symmetries. One way of breaking symmetries in a graph
is to equip the agent with some means of marking a node, i.e., making
a node locally distinguishable from all other nodes (cf.~Section~\ref{sec:model}).
For instance, the agent may have a pebble (cf.~Section~\ref{sec:summary_mapping})
-- a simple device that can be placed on a node such that the agent
recognizes the node whenever it comes back to it. A stronger model
assumes that there is a \emph{whiteboard }at each node, which allows\emph{
}the agent to leave information that it can access and modify on subsequent
visits of the node. The simple model with a single pebble is already
enough to fully break all symmetries and enable the agent to construct
a map of any graph: Starting with a map containing just the initial
node, the agent extends its map one edge at a time by traversing an
edge, marking the other end with the pebble, and then backtracking
and checking whether the now marked node was visited before and thus
already is part of the map. The model fails if there are multiple
identical and indistinguishable agents (with indistinguishable pebbles)
working in parallel. In this case not even a whiteboard at each node
is sufficient.

In directed graphs, if agents cannot look-back, an agent can still
always systematically traverse the graph assuming an upper bound on
its size is known, by trying all possible graphs and starting locations
and traversing all walks for each choice, similar to the procedure
in Section~\ref{sub:finding G^star}. However, such an exploration
algorithm generally requires exponential time. Bender et al.~\citep{BenFRSV98}
showed that if an upper bound on the size of the graph is known a
priori, an agent with a pebble can always construct a map of a directed,
strongly connected and locally oriented graph using a polynomial number
of moves. Without prior knowledge on the total number of vertices
$n$, they showed that $\Theta(\log\log n)$ pebbles are necessary
and sufficient to solve the map construction problem. 

We saw in Section~\ref{sec:base_graph} that an agent exploring any
directed graph $G$ can always construct the minimum base graph of
$G$, provided that at least an upper bound on the number of vertices
is known. When operating in an undirected graph, the same result of
course carries over. The result was originally obtained by a different
proof though, using the fact that the information encoded in the view
of a node is already contained in its view up to level $\left(n-1\right)$
\citep{Norris/95}. The resulting property is the same as in Section~\ref{sec:base_graph}.
Chalopin et al.~\citep{ChalopinDasKosowski/10} showed how to construct
the minimum-base graph of an undirected graph in polynomial time if
an upper bound on the number of vertices is known.

The initial knowledge of an agent does not necessarily have to be
about the number of vertices $n$. An important question is how much
prior information is necessary for mapping a general (undirected)
graph. It has recently been shown that, in the worst case, solving
the map construction problem for graphs of $n$ vertices and $m$
edges can require initial knowledge of $\Omega(m\log{n})$ bits~\citep{DereniowskiP/10}.
This is as much information as it takes to store the graph itself,
e.g.~if we store a list of edges.

\bibliographystyle{elsart-num-sort}
\bibliography{robots}

\begin{thebibliography}{10}
\expandafter\ifx\csname url\endcsname\relax
  \def\url#1{\texttt{#1}}\fi
\expandafter\ifx\csname urlprefix\endcsname\relax\def\urlprefix{URL }\fi

\bibitem{AwBS99}
B.~Awerbuch, M.~Betke, M.~Singh, Piecemeal graph learning by a mobile robot,
  Information and Computation 152 (1999) 155--172.

\bibitem{BenFRSV98}
M.~Bender, A.~Fernandez, D.~Ron, A.~Sahai, S.~Vadhan, The power of a pebble:
  Exploring and mapping directed graphs, in: Proceedings of the 30th ACM
  Symposium on Theory of Computing, 1998, pp. 269--287.

\bibitem{BiedlDurocherSnoeyink/09}
T.~Biedl, S.~Durocher, J.~Snoeyink, Reconstructing polygons from scanner data,
  in: Proceedings of the 20th International Symposium on Algorithms and
  Computation, 2009, pp. 862--871.

\bibitem{BiloDisserMihalakSuriVicariWidmayer/09}
D.~Bil{\`o}, Y.~Disser, M.~Mihal{\'a}k, S.~Suri, E.~Vicari, P.~Widmayer,
  Reconstructing visibility graphs with simple robots, in: Proceedings of the
  16th International Colloquium on Structural Information and Communication
  Complexity, 2009, pp. 87--99.

\bibitem{BoldiVigna/02}
P.~Boldi, S.~Vigna, Fibrations of graphs, Discrete Mathematics 243~(1--3)
  (2002) 21--66.

\bibitem{BrunnerMihalakSuriVicariWidmayer/08}
J.~Brunner, M.~Mihal{\'a}k, S.~Suri, E.~Vicari, P.~Widmayer, Simple robots in
  polygonal environments: A hierarchy, in: Proceedings of the Fourth
  International Workshop on Algorithmic Aspects of Wireless Sensor Networks,
  2008, pp. 111--124.

\bibitem{ChalopinDasDisserMihalakWidmayer/11}
J.~Chalopin, S.~Das, Y.~Disser, M.~Mihal{\'a}k, P.~Widmayer, Telling convex
  from reflex allows to map a polygon, in: Proceedings of the 28th
  International Symposium on Theoretical Aspects of Computer Science, 2011, pp.
  153--164.

\bibitem{ChalopinDasDisserMihalakWidmayer/11c}
J.~Chalopin, S.~Das, Y.~Disser, M.~Mihal\'{a}k, P.~Widmayer, Mapping simple
  polygons: How robots benefit from looking back, Algorithmica (to appear).

\bibitem{ChalopinDasKosowski/10}
J.~Chalopin, S.~Das, A.~Kosowski, Constructing a map of an anonymous graph:
  {A}pplications of universal sequences, in: Proceedings of the 14th
  international conference on Principles of distributed systems, 2010, pp.
  119--134.

\bibitem{Chvatal/75}
V.~Chv\'{a}tal, A combinatorial theorem in plane geometry, Journal of
  Combinatorial Theory (Series B) 18~(1) (1975) 39--41.

\bibitem{CoullardLubiw/91}
C.~Coullard, A.~Lubiw, Distance visibility graphs, in: Proceedings of the 7th
  Annual Symposium on Computational Geometry, 1991, pp. 289--296.

\bibitem{DereniowskiP/10}
D.~Dereniowski, A.~Pelc, Drawing maps with advice, in: Proceedings of the 24th
  International Symposium on Distributed Computing, 2010, pp. 328--342.

\bibitem{DisserPhD/11}
Y.~Disser, Mapping polygons, Ph.D. thesis, ETH Zurich (2011).

\bibitem{DisserMihalakWidmayer/10b}
Y.~Disser, M.~Mihal\'{a}k, P.~Widmayer, Reconstruction of a polygon from angles
  without prior knowledge of the size, Tech. Rep. 700, ETH Z{\"u}rich,
  Institute of Theoretical Computer Science (November 2010).

\bibitem{DisserMihalakWidmayer/11}
Y.~Disser, M.~Mihal{\'a}k, P.~Widmayer, A polygon is determined by its angles,
  Computational Geometry: Theory and Applications 44 (2011) 418--426.

\bibitem{Everett/90}
H.~Everett, Visibility graph recognition, Ph.D. thesis, University of Toronto,
  Department of Computer Science (January 1990).

\bibitem{Fisk/78}
S.~Fisk, A short proof of {C}hvatal's watchman theorem, Journal of
  Combinatorial Theory (Series B) 24~(3) (1978) 374.

\bibitem{FraIPPP04}
P.~Fraigniaud, D.~Ilcinkas, G.~Peer, A.~Pelc, D.~Peleg, Graph exploration by a
  finite automaton, in: Proceedings of the 29th Symposium on Mathematical
  Foundations of Computer Science, 2004, pp. 451--462.

\bibitem{GanguliCortesBullo/06}
A.~Ganguli, J.~Cort{\'e}s, F.~Bullo, Distributed deployment of asynchronous
  guards in art galleries, in: Proceedings of the 2006 American Control
  Conference, 2006, pp. 1416--1421.

\bibitem{GfellerMihalakSuriVicariWidmayer/07}
B.~Gfeller, M.~Mihalak, S.~Suri, E.~Vicari, P.~Widmayer, Counting targets with
  mobile sensors in an unknown environment, in: Proceedings of the 3rd
  International Workshop on Algorithmic Aspects of Wireless Sensor Networks,
  2007, pp. 32--45.

\bibitem{Ghosh/07}
S.~K. Ghosh, Visibility Algorithms in the Plane, Cambridge University Press,
  2007.

\bibitem{GhoshGoswami/09}
S.~K. Ghosh, P.~P. Goswami, Unsolved problems in visibility graph theory, in:
  Proceedings of the India-Taiwan Conference on Discrete Mathematics, 2009, pp.
  44--54.

\bibitem{JacksonWismath/02}
L.~Jackson, S.~Wismath, Orthogonal polygon reconstruction from stabbing
  information, Computational Geometry 23~(1) (2002) 69--83.

\bibitem{KomuravelliMihalak/09}
A.~Komuravelli, M.~Mihal{\'a}k, Exploring polygonal environments by simple
  robots with faulty combinatorial vision, in: Proceedings of the 11th
  International Symposium on Stabilization, Safety, and Security of Distributed
  Systems, 2009, pp. 458--471.

\bibitem{Norris/95}
N.~Norris, Universal covers of graphs: isomorphism to depth $n-1$ implies
  isomorphism to all depths, Discrete Applied Mathematics 56~(1) (1995) 61--74.

\bibitem{ORourke/88}
J.~O'Rourke, Uniqueness of orthogonal connect-the-dots, in: G.~Toussaint (ed.),
  Computational Morphology, North-Holland, 1988, pp. 97--104.

\bibitem{ORourkeStreinu/98}
J.~O'Rourke, I.~Streinu, The vertex-edge visibility graph of a polygon,
  Computational Geometry 10~(2) (1998) 105--120.

\bibitem{PanPe99}
P.~Panaite, A.~Pelc, Exploring unknown undirected graphs, Journal of Algorithms
  33 (1999) 281--295.

\bibitem{Rappaport/86}
D.~Rappaport, On the complexity of computing orthogonal polygons from a set of
  points, Tech. Rep. SOCS-86.9, McGill University, Montreal, Canada (1986).

\bibitem{Reingold/05}
O.~Reingold, Undirected st-connectivity in log-space, in: Proceedings of the
  37th Annual ACM Symposium on Theory of Computing, 2005, pp. 376--385.

\bibitem{SidleskyBarequetGotsman/06}
A.~Sidlesky, G.~Barequet, C.~Gotsman, Polygon reconstruction from line
  cross-sections, in: Proceedings of the 18th Annual Canadian Conference on
  Computational Geometry, 2006, pp. 81--84.

\bibitem{Snoeyink/99}
J.~Snoeyink, Cross-ratios and angles determine a polygon, Discrete and
  Computational Geometry 22~(4) (1999) 619--631.

\bibitem{SuriVicariWidmayer/08}
S.~Suri, E.~Vicari, P.~Widmayer, Simple robots with minimal sensing: From local
  visibility to global geometry, International Journal of Robotics Research
  27~(9) (2008) 1055--1067.

\bibitem{YamashitaKameda/96}
M.~Yamashita, T.~Kameda, Computing on anonymous networks: Part {I} --
  characterizing the solvable cases, IEEE Transactions on Parallel and
  Distributed Systems 7~(1) (1996) 69--89.

\end{thebibliography}

\end{document}